\documentclass[fleqn,usenatbib]{mnras}
    \usepackage{newtxtext,newtxmath}
    \usepackage[T1]{fontenc}
    \usepackage{booktabs}
    \usepackage{multirow} 
    \usepackage{makecell}
    \usepackage{graphicx}
    \usepackage{longtable}
    \usepackage{amsmath}

    \graphicspath{{./}}

    \newcommand{\EWSiXIVKaMEGLaraDI}{{$1.06\pm 0.20$}}          
    \newcommand{\EWSXVIKaMEGLaraDI}{{$1.07^{+0.32}_{-0.27}$}}   
    \newcommand{\EWSiXIVKaLETGLaraDI}{{$3.49^{+1.28}_{-1.30}$}} 
    \newcommand{\EWSXVIKaLETGLaraDI}{{$4.87^{+1.79}_{-1.75}$}}  

        \newcommand{\NHSiXIVKaMEGLaraDI}{{$8.7\pm{1.6}\times10^{15}$}}          
    \newcommand{\NHSXVIKaMEGLaraDI}{{$1.50^{+0.44}_{-0.38}\times10^{16}$}}     
    \newcommand{\NHSiXIVKaLETGLaraDI}{{$2.85^{+1.05}_{-1.06}\times10^{16}$}} 
    \newcommand{\NHSXVIKaLETGLaraDI}{{$6.80^{+2.51}_{-2.44}\times10^{16}$}}  

    \newcommand{\PosNVIKb}{{24.898}}    

    \newcommand{\PosNVIIKa}{{24.781}}   

    \newcommand{\PosOIIKb}{{22.04}}    

    \newcommand{\PosOVIKa}{{22.034}}    

    \newcommand{\PosOVIIKa}{{21.601}}   
    \newcommand{\PosOVIIKb}{{18.627}}   
    \newcommand{\PosOVIIKg}{{17.768}}   

    \newcommand{\PosOVIIIKa}{{18.969}}  

    \newcommand{\PosSiXIVKa}{{6.182}}   

    \newcommand{\PosSXVIKa}{{4.729}}    

    \newcommand{\PosNeIXKa}{{13.447}}   

\newcommand{\delchiMEGWARM}{{2}}
\newcommand{\delchiMEGWARMHOT}{{61}}
\newcommand{\delchiMEGHOT}{{57}}
\newcommand{\delchiLETGWARM}{{36}}
\newcommand{\delchiLETGWARMHOT}{{10}}
\newcommand{\delchiLETGHOT}{{31}}

    \newcommand{\TempMEGWARM}{{5.39$_{-0.07}^{+0.13}$}}         
    \newcommand{\NHMEGWARM}{{18.08$_{-0.40}^{+0.31}$}}                   
    \newcommand{\TempMEGWARMHOT}{{6.19$_{-0.08}^{+0.06}$}}      
    \newcommand{\NHMEGWARMHOT}{{19.12$_{-0.10}^{+0.08}$}}       
    \newcommand{\TempMEGSUPERHOT}{{7.50$\pm{0.03}$}}      
    \newcommand{\NHMEGSUPERHOT}{{21.30$_{-0.06}^{+0.05}$}}       
    \newcommand{\TempLETGWARM}{{5.64$\pm{0.06}$}}         
    \newcommand{\NHLETGWARM}{{19.76$_{-0.11}^{+0.09}$}}                   
    \newcommand{\TempLETGWARMHOT}{{6.56$\pm{0.12}$}}      
    \newcommand{\NHLETGWARMHOT}{{19.56$_{-0.27}^{+0.21}$}}       
    \newcommand{\TempLETGSUPERHOT}{{7.50$\pm{0.05}$}}      
    \newcommand{\NHLETGSUPERHOT}{{21.75$_{-0.14}^{+0.08}$}}       

    \newcommand{\EWNVIIKaMEGWARMHOT}{{$3.26$}}
    
    \newcommand{\EWNVIKbMEGWARM}{{$1.63$}}
    \newcommand{\EWNVIKbMEGWARMHOT}{{$1.02$}}
    
    \newcommand{\EWNeIXKaLETGWARMHOT}{{$2.57$}}
    \newcommand{\EWNeIXKaMEGWARMHOT}{{$1.32$}}
    \newcommand{\EWOVIIIKaLETGHOT}{{$8.51$}}
    \newcommand{\EWOVIIIKaLETGWARMHOT}{{$6.42$}}
    \newcommand{\EWOVIIIKaMEGHOT}{{$4.27$}}
    \newcommand{\EWOVIIIKaMEGWARMHOT}{{$1.83$}}
    
    \newcommand{\EWOVIIKaLETGWARM}{{$15.81$}}
    \newcommand{\EWOVIIKaLETGWARMHOT}{{$3.87$}}
    \newcommand{\EWOVIIKaMEGWARMHOT}{{$11.63$}}
    
    \newcommand{\EWOVIIKbMEGWARMHOT}{{$3.07$}}

    \newcommand{\EWOVIIKgMEGWARMHOT}{{$1.18$}}
    
    \newcommand{\EWOVIKaLETGWARM}{{$3.75$}}
    
    \newcommand{\EWSXVIKaLETGHOT}{{$5.05$}}
    \newcommand{\EWSXVIKaMEGHOT}{{$0.84$}}

    \newcommand{\EWSiXIVKaLETGHOT}{{$3.85$}}
    \newcommand{\EWSiXIVKaMEGHOT}{{$0.93$}}
    \newcommand{\NHNVIIMEGWARMHOT}{{$1.64\times10^{15}$}}
    
    \newcommand{\NHNVIMEGWARM}{{$2.57\times10^{15}$}}
    \newcommand{\NHNVIMEGWARMHOT}{{$1.35\times10^{15}$}}
    \newcommand{\NHNeIXLETGWARMHOT}{{$2.49\times10^{15}$}}
    \newcommand{\NHNeIXMEGWARMHOT}{{$1.24\times10^{15}$}}
    \newcommand{\NHOVIIILETGHOT}{{$7.03\times10^{15}$}}
    \newcommand{\NHOVIIILETGWARMHOT}{{$6.09\times10^{15}$}}
    \newcommand{\NHOVIIIMEGHOT}{{$3.48\times10^{15}$}}
    \newcommand{\NHOVIIIMEGWARMHOT}{{$1.51\times10^{15}$}}
    \newcommand{\NHOVIILETGWARM}{{$3.95\times10^{16}$}}
    \newcommand{\NHOVIILETGWARMHOT}{{$1.5\times10^{15}$}}
    \newcommand{\NHOVIIMEGWARMHOT}{{$8.22\times10^{15}$}}
    \newcommand{\NHOVILETGWARM}{{$2.42\times10^{15}$}}
    
    \newcommand{\NHSXVILETGHOT}{{$7.83\times10^{16}$}}
    \newcommand{\NHSXVIMEGHOT}{{$1.09\times10^{16}$}}
    
    \newcommand{\NHSiXIVLETGHOT}{{$3.12\times10^{16}$}}
    \newcommand{\NHSiXIVMEGHOT}{{$6.99\times10^{15}$}}
    \newcommand{\PosNVIIKaMEGWARMHOT}{{$24.788$}}
    
    \newcommand{\PosNVIKbMEGWARM}{{$24.898$}}
    \newcommand{\PosNVIKbMEGWARMHOT}{{$24.905$}}
    
    \newcommand{\PosNeIXKaLETGWARMHOT}{{$13.448$}}
    \newcommand{\PosNeIXKaMEGWARMHOT}{{$13.451$}}
    \newcommand{\PosOVIIIKaLETGHOT}{{$18.984$}}
    \newcommand{\PosOVIIIKaLETGWARMHOT}{{$18.970$}}
    \newcommand{\PosOVIIIKaMEGHOT}{{$18.973$}}
    \newcommand{\PosOVIIIKaMEGWARMHOT}{{$18.974$}}
    
    \newcommand{\PosOVIIKaLETGWARM}{{$21.603$}}
    \newcommand{\PosOVIIKaLETGWARMHOT}{{$21.602$}}
    \newcommand{\PosOVIIKaMEGWARMHOT}{{$21.607$}}
    
    \newcommand{\PosOVIIKbMEGWARMHOT}{{$18.632$}}

    \newcommand{\PosOVIIKgMEGWARMHOT}{{$17.773$}}
    
    \newcommand{\PosOVIKaLETGWARM}{{$22.037$}}
    
    \newcommand{\PosSXVIKaLETGHOT}{{$4.733$}}
    \newcommand{\PosSXVIKaMEGHOT}{{$4.730$}}

    \newcommand{\PosSiXIVKaLETGHOT}{{$6.187$}}
    \newcommand{\PosSiXIVKaMEGHOT}{{$6.184$}}

    \newcommand{\XSPEC}{{\fontfamily{qcr}\selectfont XSPEC}}        

    \newcommand{\pow}{{\fontfamily{qcr}\selectfont powerlaw}}       
    \newcommand{\tbabs}{{\fontfamily{qcr}\selectfont Tbabs}}        
    \newcommand{\bbody}{{\fontfamily{qcr}\selectfont bbody}}        
    \newcommand{\agaus}{{\fontfamily{qcr}\selectfont agauss}}       
    \newcommand{\Ka}{{\fontfamily{qcr}\selectfont K$\alpha$}}       
    \newcommand{\Kb}{{\fontfamily{qcr}\selectfont K$\beta$}}        
    \newcommand{\Kg}{{\fontfamily{qcr}\selectfont K$\gamma$}}       

    \newcommand{\Chandra}{{\fontfamily{qcr}\selectfont Chandra}}    
    \newcommand{\XMMN}{{\fontfamily{qcr}\selectfont XMM-Newton}}    

    \newcommand{\MEG}{{\fontfamily{qcr}\selectfont MEG}}            
    \newcommand{\LETG}{{\fontfamily{qcr}\selectfont LETG}}          

    \newcommand{\ACISS}{{\fontfamily{qcr}\selectfont ACIS-S}}                   

    \newcommand{\warm}{{\fontfamily{qcr}\selectfont warm}}                      
    \newcommand{\warmhot}{{\fontfamily{qcr}\selectfont warm-hot}}               
    \newcommand{\hot}{{\fontfamily{qcr}\selectfont hot}}                        
       
    \newcommand{\MEGWARM}{{\fontfamily{qcr}\selectfont MEG-WARM}}               
    \newcommand{\MEGWARMHOT}{{\fontfamily{qcr}\selectfont MEG-WARM-HOT}}        
    \newcommand{\MEGHOT}{{\fontfamily{qcr}\selectfont MEG-HOT}}                 
    \newcommand{\MEGSUPERHOT}{{\fontfamily{qcr}\selectfont MEG-HOT}}            

    \newcommand{\LETGWARM}{{\fontfamily{qcr}\selectfont LETG-WARM}}             
    \newcommand{\LETGWARMHOT}{{\fontfamily{qcr}\selectfont LETG-WARM-HOT}}      
    \newcommand{\LETGHOT}{{\fontfamily{qcr}\selectfont LETG-HOT}}               
    \newcommand{\LETGSUPERHOT}{{\fontfamily{qcr}\selectfont LETG-HOT}}          

    \newcommand{\PHASE}{{\fontfamily{qcr}\selectfont PHASE}}                        

    \newcommand{\kms}{$\mathrm{km} \ \mathrm{s}^{-1}$}  
    \newcommand{\logT}{$\log(T/\mathrm{K})$}            

    \newcommand{\delchi}{{\fontfamily{qcr}\selectfont $\Delta\chi^2$}}                      

    \newcommand{\chidof}{{\fontfamily{qcr}\selectfont $\chi^2/\mathrm{d.o.f}$}}             

    \newcommand{\logNH}{{\fontfamily{qcr}\selectfont $\log (\mathrm{N}_\mathrm{H}$}/cm$^{-2}$)}         

    \newcommand{\Oxygen}{{\fontfamily{qcr}\selectfont O}}                               
    \newcommand{\Ne}{{\fontfamily{qcr}\selectfont Ne}}                                  
    \newcommand{\Si}{{\fontfamily{qcr}\selectfont Si}}                                  
    \newcommand{\Sulfur}{{\fontfamily{qcr}\selectfont S}}                               
    \newcommand{\Fe}{{\fontfamily{qcr}\selectfont Fe}}                                  

    \newcommand{\NVI}{{\fontfamily{qcr}\selectfont N$_{\mathrm{VI}}$}}
    \newcommand{\NVII}{{\fontfamily{qcr}\selectfont N$_{\mathrm{VII}}$}}

    \newcommand{\NeIX}{{\fontfamily{qcr}\selectfont Ne$_{\mathrm{IX}}$}}                
    \newcommand{\NeX}{{\fontfamily{qcr}\selectfont Ne$_{\mathrm{X}}$}}                  
    \newcommand{\NeXI}{{\fontfamily{qcr}\selectfont Ne$_{\mathrm{XI}}$}}                

    \newcommand{\OII}{{\fontfamily{qcr}\selectfont O$_{\mathrm{II}}$}}                  
    \newcommand{\OVI}{{\fontfamily{qcr}\selectfont O$_{\mathrm{VI}}$}}                  
    \newcommand{\OVII}{{\fontfamily{qcr}\selectfont O$_{\mathrm{VII}}$}}                
    \newcommand{\OVIII}{{\fontfamily{qcr}\selectfont O$_{\mathrm{VIII}}$}}              

    \newcommand{\SXVI}{{\fontfamily{qcr}\selectfont S$_{\mathrm{XVI}}$}}                

    \newcommand{\SiXIV}{{\fontfamily{qcr}\selectfont Si$_{\mathrm{XIV}}$}}              

    \title[Sub-solar Fe/O in the Hot Milky Way CGM]{A Sub-solar \Fe/\Oxygen, \logT\ $\sim$ 7.5 Gas Component Permeating the Milky Way's CGM}

    \author[Lara-DI et al.]{Armando Lara-DI,$^{1}$
    Yair Krongold,$^{1}$
    Smita Mathur,$^{2,3}$
    Sanskriti Das,$^{4}$
    Anjali Gupta,$^{2}$
    O. Segura Montero,$^{1}$
    \\
    $^{1}$Instituto de Astronomia, Universidad Nacional Autonoma de México, 04510 Mexico City, Mexico\\
    $^{2}$Astronomy Department, The Ohio State University, Columbus, OH 43210\\
    $^{3}$Center for Cosmology and Astro-Particle Physics, The Ohio State University, Columbus, OH 43220\\
    $^{4}$Kavli Institute for Particle Astrophysics \& Cosmology, Stanford University, 452 Lomita Mall, Stanford, CA 94305, USA
    }

    \date{Accepted XXX. Received YYY; in original form ZZZ}

    \pubyear{2024}
\begin{document}
    \label{firstpage}
    \pagerange{\pageref{firstpage}--\pageref{lastpage}}
    \maketitle


    \begin{abstract}
            
        {Our study focuses on characterizing the highly ionized gas within the Milky Way's (MW) Circumgalactic Medium (CGM) that gives rise to ionic transitions in the X-ray band 2 - 25 \AA. Utilizing stacked \Chandra/\ACISS\ \MEG\ and \LETG\ spectra toward QSO sightlines, we employ the self-consistent hybrid ionization code PHASE to model our data. The stacked spectra are optimally described by three distinct gas phase components: a \warm\ (\logT\ $\sim$ 5.5), \warmhot\ (\logT\ $\sim 6$), and \hot\ (\logT\ $\sim$ 7.5) components. These findings confirm the presence of the \hot\ component in the MW's CGM indicating its coexistence with a \warm\ and a \warmhot\ gas phases. We find this \hot\ component to be homogeneous in temperature but inhomogeneous in column density. The gas in the \hot\ component requires over-abundances relative to solar to be consistent with the Dispersion Measure (DM) from the Galactic halo reported in the literature. {For the hot phase we estimated a DM = $55.1^{+29.9}_{-23.7}$ pc cm$^{-3}$}. We conclude that this phase is either enriched in Oxygen, Silicon, and Sulfur, or has metallicity {over 6} times solar value, or a combination of both. We do not detect Fe L-shell absorption lines, implying O/Fe $\geq$ 4. The non-solar abundance ratios found in the super-virial gas component in the Galactic halo suggest that this phase arises from Galactic feedback.}

    \end{abstract}


%
%
    \section{Introduction} \label{sec:intro}

        The circumgalactic medium (CGM) refers to the gaseous component located beyond the galactic disc and within the galactic virial radius. The CGM is a mixed medium with complex structures like filaments, bubbles, and multiphase regions. It comprises ionized and neutral gas with different temperatures and densities (e.g., \citealp{Tumlinson2017}; \citealp{Mathur2022}). 

        The CGM is thought to play a crucial role in regulating the exchange of matter and energy between the galactic disc and its environment (e.g., \citealp{Keres2005}, \citealp{Zheng2015}). Numerical simulations have shown that shock-heating processes might have heated and ionized the gas during galaxy formation, preventing it from falling into the galactic disc. On the other hand, feedback processes such as supernovas and galactic winds could have expelled large amounts of material into the CGM (e.g., \citealp{Stinson2012}). Processes such as the infall of material (commonly pristine gas) from the intergalactic medium towards the galactic disc and the expulsion of metals formed in stars from the disc into the surrounding region make the CGM an important clue to understanding galactic formation and evolution (e.g., \citealp{Tumlinson2017}; \citealp{Li2018}). These studies suggest that the CGM is a large reservoir of gas that can fuel future star formation within the galaxy and the place where the missing baryons and metals could reside.
        
        In the local universe, at galactic scales, the number of baryons observed in galaxies with luminosity less or near Schechter L$^{\star}$ lies near half under the amount of baryonic mass predicted by the nucleosynthesis in the Big Bang theory and inferred from the density fluctuations of the cosmic microwave background. For galaxies less massive, more mass is missing (e.g., \citealp{Kirkman2003} and references therein; \citealp{PlanckCollaboration2016}, \citealp{McGaugh2010}). Coupled with the Missing Baryon Problem, there is also The Missing Metal Problem at galactic scales. The number of metals expected from the stars observed and the star formation history of galaxies is about two times larger than the number of metals we can observe (\citealp{Peeples2014}). 
        
        Theoretical studies suggest that the missing baryons and missing metals in galaxies might reside in the CGM; however, its diffuse nature makes it difficult to study this material (e.g., \citealp{Feldmann2013}; \citealp{Mathur2021}; \citealp{Mathur2023}). 
        
        Because of our unique point of reference, studying the CGM of the Milky Way is much easier than studying it in external galaxies. A common way to detect the CGM of the Milky Way is by studying it in absorption against bright background sources, such as quasars (e.g., \citealp{Gupta2012}, \citealp{Mathur2022}).
            
        The CGM is expected to be close to the galaxy's virial temperature. For a Milky Way-like galaxy, this temperature is about \logT\ $\sim$ 6. We will call the gas at this high temperature the \warmhot\ component. For the Milky Way, this gas phase has been studied in emission and absorption using UV and X-ray spectra (e.g., \citealp{Wang2005}; \citealp{Gupta2012}; \citealp{Das2019}). 
        
        \citet{Gupta2012} detected in absorption towards extragalactic sight lines \OVII\ and \OVIII\ at $z=0$. Using and combining their results with emission results from \citealp{Henley2010}, they derive that the \warmhot\ phase of the Milky Way CGM is a massive component of about $\log$($M$/M$_\odot$) $\sim$ 10, which extends over a vast region around the galactic disc. Their results suggest that it is in this component where the missing baryons and metals could reside.
        
        In recent studies, \citet{Das2019}, \citet{Das2021}, and  detected a hotter gas component in the Milky Way CGM for the first time. They found a gas phase with super-virial temperature at \logT\ $\gtrsim 7$. Hereafter, we will refer to this component as \hot. Using deep \XMMN\ RGS observations of the blazar 1ES 1553+113, \citet{Das2019} detected in absorption \NeX\ \Ka\ associated with this \hot\ component. Later on, in 2021 \citet{Das2021} also found \SiXIV\ \Ka\ and \NeX\ \Ka\ in the line of sight towards the blazar Mkn 421. Their results in these two lines of sight indicate that the CGM is a multiphase system with a \warm\ (\logT\ $\sim$ 5.5), \warmhot\ (\logT\ $\sim$ 6), and \hot\ (\logT\ $\sim$ 7.5) components.
        
        Focused on these gas components, and using \Chandra\ observations towards 47 different sightlines and the stacking technique, \citet{LaraDI2023} (hereafter LDI-2023a) were able to detect \SiXIV\ \Ka\ and for the first time \SXVI\ \Ka\ at $z=0$ in absorption. Their discovery confirms the presence of a \hot\ component in the Milky Way CGM, suggesting it is a widespread component throughout the entire CGM. The presence of the hot component was also confirmed by \citet{McClain2023} in the sightline toward NGC\,3783.
    
        The newly discovered \hot\ component was also detected in emission (e.g., \citealp{Das2019apj887257}; \citealp{Bluem2022}; \citealp{Gupta2021}; \citealp{Bhattacharyya2023}; \citealp{Gupta2023}). 
        \citet{Bhattacharyya2023} studied in emission the \hot\ phase of the CGM around the Mkn 421 sightline. Their study complements the work done by \citet{Das2021} in the line of sight towards Mkn 421, showing that the emitting gas has higher density, possibly coming from regions close to the Galactic disc. On the other hand, the absorption measurements arise from low densities extending to the virial radius. They found a scatter in the temperature in both the \warmhot\ component and the \hot\ component; this contributes to understanding the CGM as a multiphase system.
        
        Despite these recent efforts, characterizing the CGM remains challenging. Information about its geometry, homogeneity, and how the different gas phases expand through the entire CGM Milky Way remains to be determined. The super-virial \hot\ component is not expected from theoretical studies, and its finding opens new questions crucial to better understanding galactic formation and evolution.
        
        This paper focuses on absorption studies of the Milky Way CGM. We use the LDI-2023a data sample to characterize the CGM by identifying the gas phases from which these and other absorption lines come. LDI-2023a focused on the spectral range 4 - 8 \AA; now, we will focus on the study of the spectral range from 2 to 25 \AA\ using a self-consistent ionization model (\PHASE, \citealp{Krongold2003}) to fit the data.
        We structure our paper as follows. In Section 2, we present the sample selection. In Section 3, we describe the data analysis. In Section 4, we show our results, and in Section 5, we discuss their implications. 
              
%
%

    \section{Data Sample} \label{sec:datasample}

        This paper uses the same data sample and stacked spectra used in LDI-2023a. The data sample consists of 47 different sight lines to QSOs, Seyfert-1, and Blazars from Chandra X-ray Observatory public observations. This sample excludes changing look Active Galactic Nuclei (AGN)\footnote{Mkn 590}, which sometimes act like a Seyfert-1 and sometimes as Syfert-2 objects. We also exclude long observations with very high S/N\footnote{3C 273 and PKS 2155-304} that would dominate the stacked spectra in our analysis, along with ES1553+113 and Mkn 421 sightlines used by \citet{Das2019} and \citet{Das2021}. Finally, NGC 4051 ($z=0.002$), which presents at $z=0$ clear contribution of narrow absorption lines due to its Warm Absorber (WA), was also excluded (e.g., \citealp{Krongold2007}). In this way, we avoid intrinsic absorption lines of the WA contaminating our analysis. The sightlines chosen have high Galactic latitude, meaning our data has a small cross-section with the Milky Way's Interstellar Medium (ISM) (see LDI-2023a for more details).

        The stacked spectra with \MEG\ observations comprise 46 (10.96 Ms) of the 47 sightlines, while \LETG\ has nine (1.09 Ms). Except for one sightline (TON S 180), all the sightlines covered by \LETG\ are contained within \MEG. The contribution of \LETG\ sightlines to the overall exposure time of \MEG\ amounts to about 10$\%$. It is important to note that \MEG\ data is dominated by sources with WA, whereas \LETG\ is not.
    
        We study the spectral range 2 - 25 \AA, with a signal-to-noise (S/N) ratio of 2181 for \MEG\ and 651 for \LETG. In LDI-2023a, the complete list of sightlines, the Aitoff projection of these targets, and the complete table with the list of individual observations are included. See also LDI-2023a for details on how the data was reprocessed and stacked.

%
%

    \section{Analysis \label{sec:analysis}}


    \subsection{Fitting the Continuum}

        To analyze \MEG\ and \LETG\ stacked spectra, we used the spectral fitting software \XSPEC\ (v12.13.0) with $\chi^2$ statistics. Errors presented in this paper correspond to 1$\sigma$ level.
    
        We first modeled the continuum of each spectrum from 2.0 to 25.0 \AA. We used a model comprising a power law (\pow), a black body (\bbody) component, an ISM absorption component (\tbabs), and as many Gaussian lines (\agaus) as required to account for AGN intrinsic absorption or emission features (i.e., lines produced intrinsically in the sources and not arising from the Milky Way's CGM). 
        

    \subsection{\PHASE}

    To model the absorption lines on the spectra arising from the gas components at $z=0$ of the Milky's Way CGM, we used the hybrid (photo + collision) ionization code \PHASE\ \citep{Krongold2003}. This code allows the analysis of X-ray spectra by generating a synthetic spectrum that fits the data and gives the gas phase component that best describes it. \PHASE\ has as free parameters the temperature of the gas ($T$), the hydrogen column density ($N_H$), the redshift ($z$), the micro-turbulent velocity of the gas ($v$), the ionization parameter ($U$), and the abundances of the elements, which are set to solar by default. 
 
    The CGM is expected to have densities for which photoionization from the metagalactic background should be negligible. Therefore, in spectroscopic modeling we fixed the ionization parameter of the gas to low values ($\log\mathrm{U} \sim -4$). Since our study focuses on the Milky Way's CGM, we set the redshift $z\approx0$. The width of the lines was set to be less than the resolution element of the instrument. Therefore, our model is constrained to fit narrow lines. In particular, $v \approx 10$ \kms.
   
    This way, we added a \PHASE\ component to the continuum model. This first component was set to fit the virial component (\warmhot) by only considering spectral ranges where \NeIX\ and \OVII\ appear and letting fit the model to the best \logT\ and \logNH\ for this component. Once we had the best temperature for this component, we extended our analysis to cover all spectral ranges and introduced an additional hotter component (\hot) with a higher temperature. Finally, we included a warmer (\warm) component with a lower temperature and fitted the three components simultaneously. However, it is important to note that our results do not depend on the order in which the \PHASE\ components are added to the continuum.
   
    The final statistic for the model fitting the continuum on \MEG\ was \chidof\ = 6600.48/4599, and \chidof\ = 1606.56/1837 on \LETG.

        
        \subsubsection{Abundances}

        In the X-ray band, we do not have a diagnostic for Hydrogen, so we cannot constrain the absolute metallicity of the CGM. However, we can study the relative abundances of different elements. To do this, we allow each element's abundance to vary independently when necessary. 

%
%

    \section{Results} \label{sec:results}    

\begin{figure*}
        \centering
        \begin{tabular}{ccc}
            \includegraphics[width=0.33\textwidth]{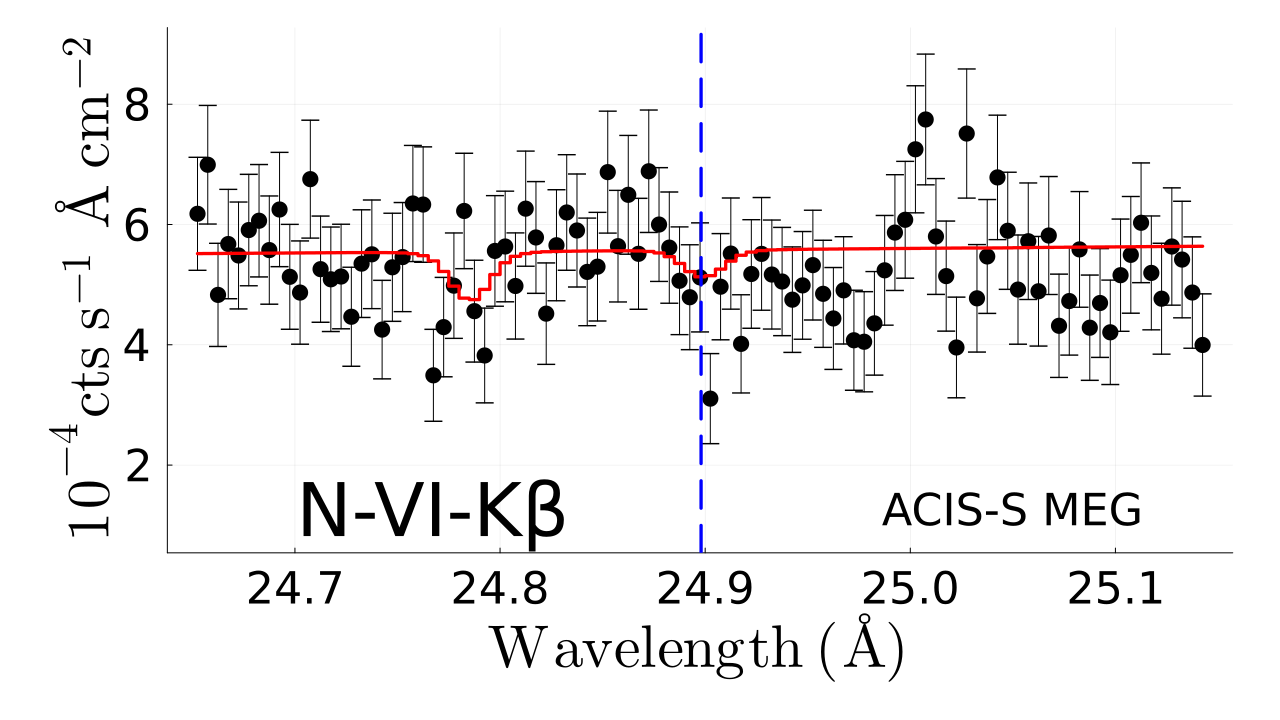}   & 
            \includegraphics[width=0.33\textwidth]{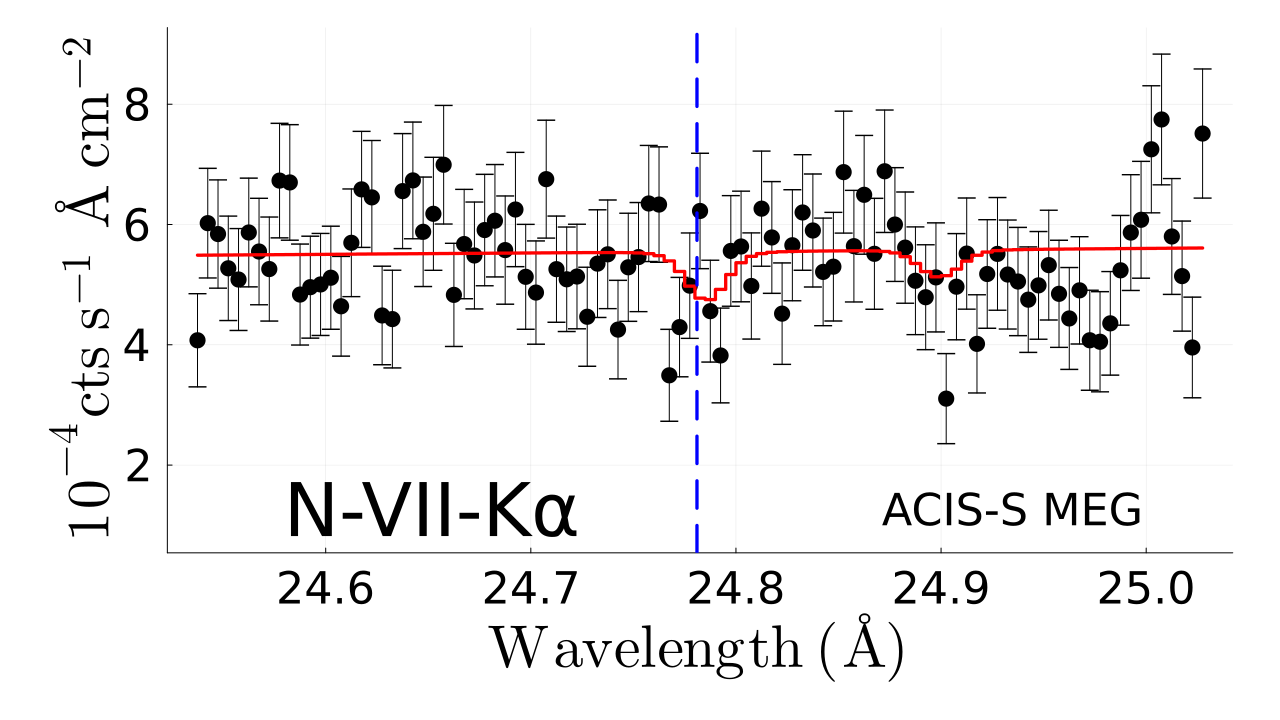}   & 
            \includegraphics[width=0.33\textwidth]{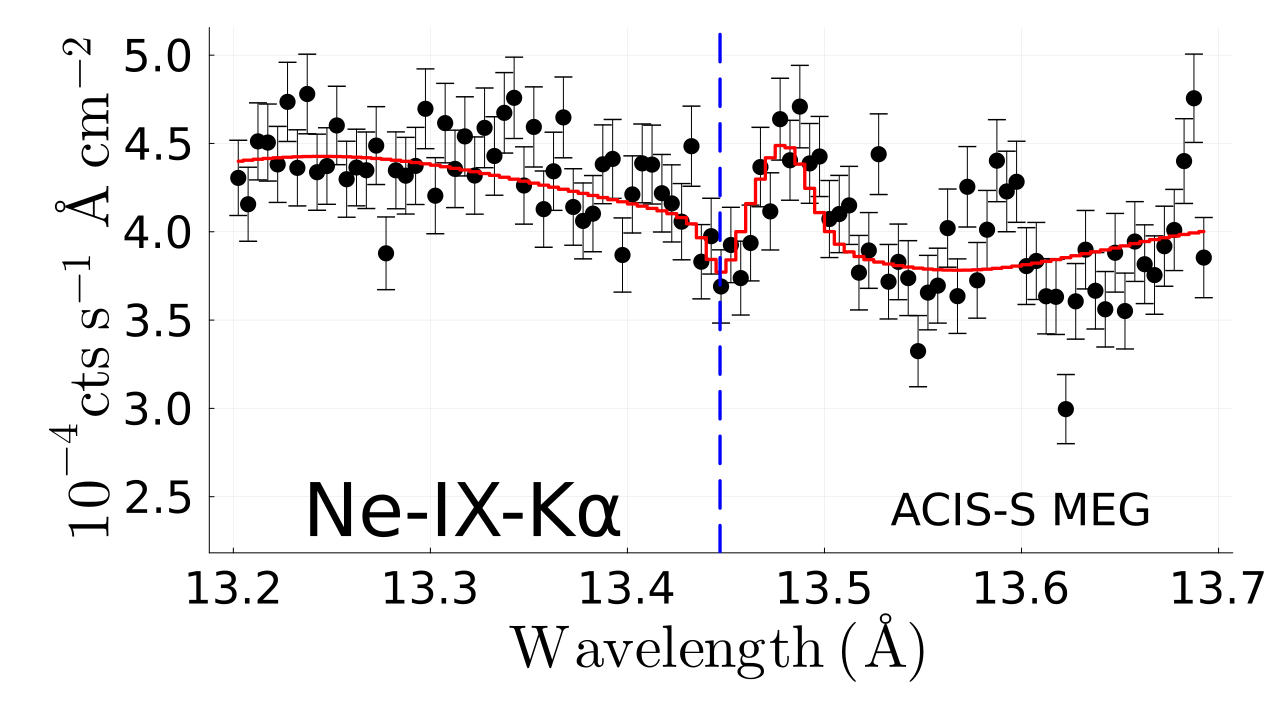}    \\
            \includegraphics[width=0.33\textwidth]{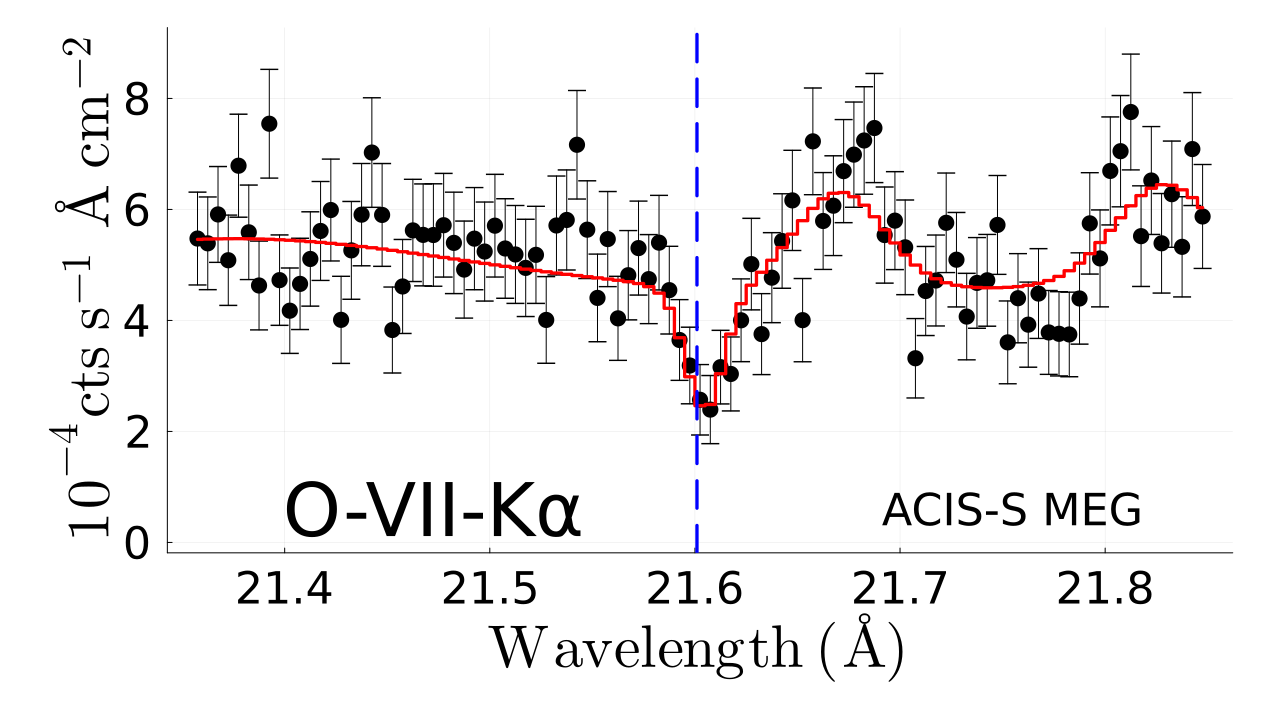}   &
            \includegraphics[width=0.33\textwidth]{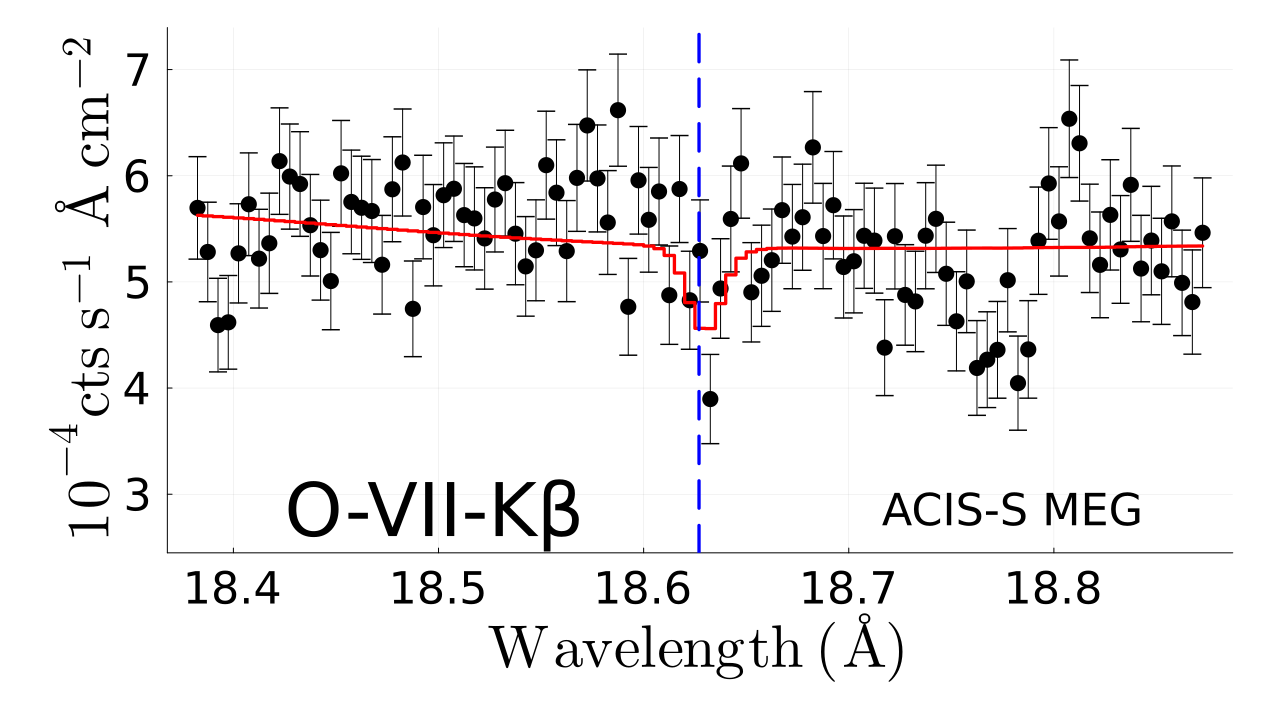}   &
            \includegraphics[width=0.33\textwidth]{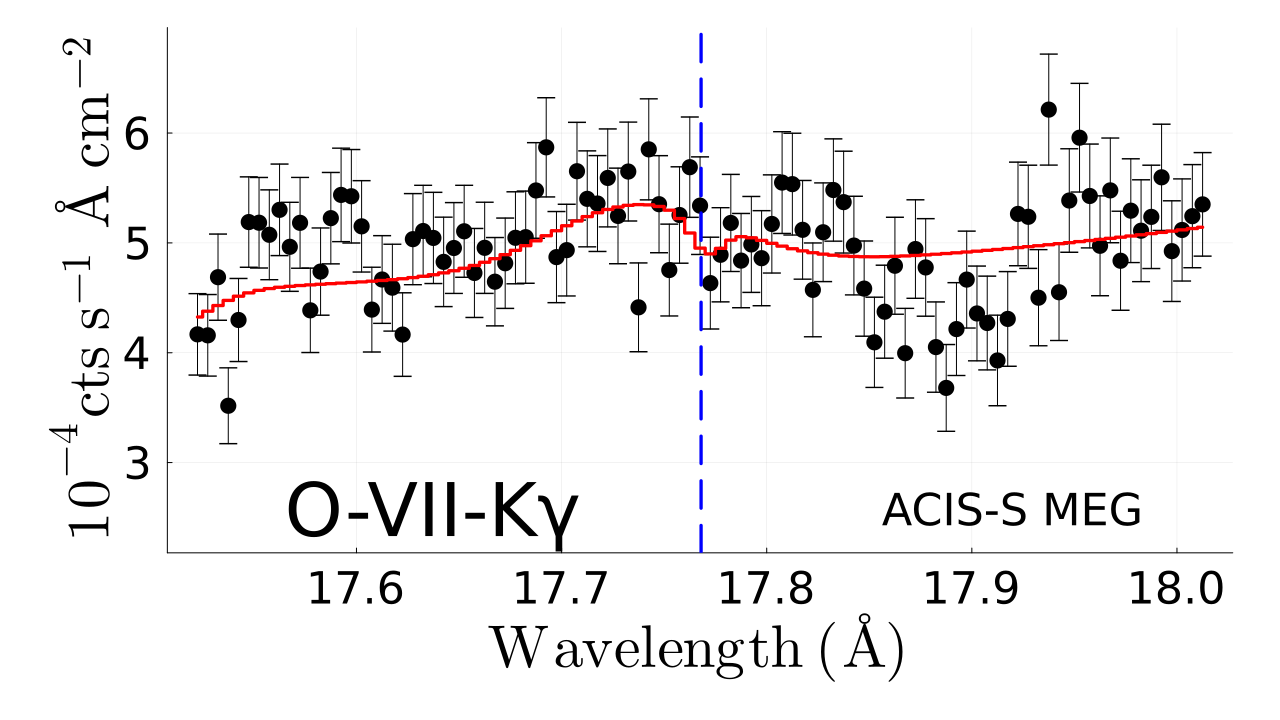}   \\
            \includegraphics[width=0.33\textwidth]{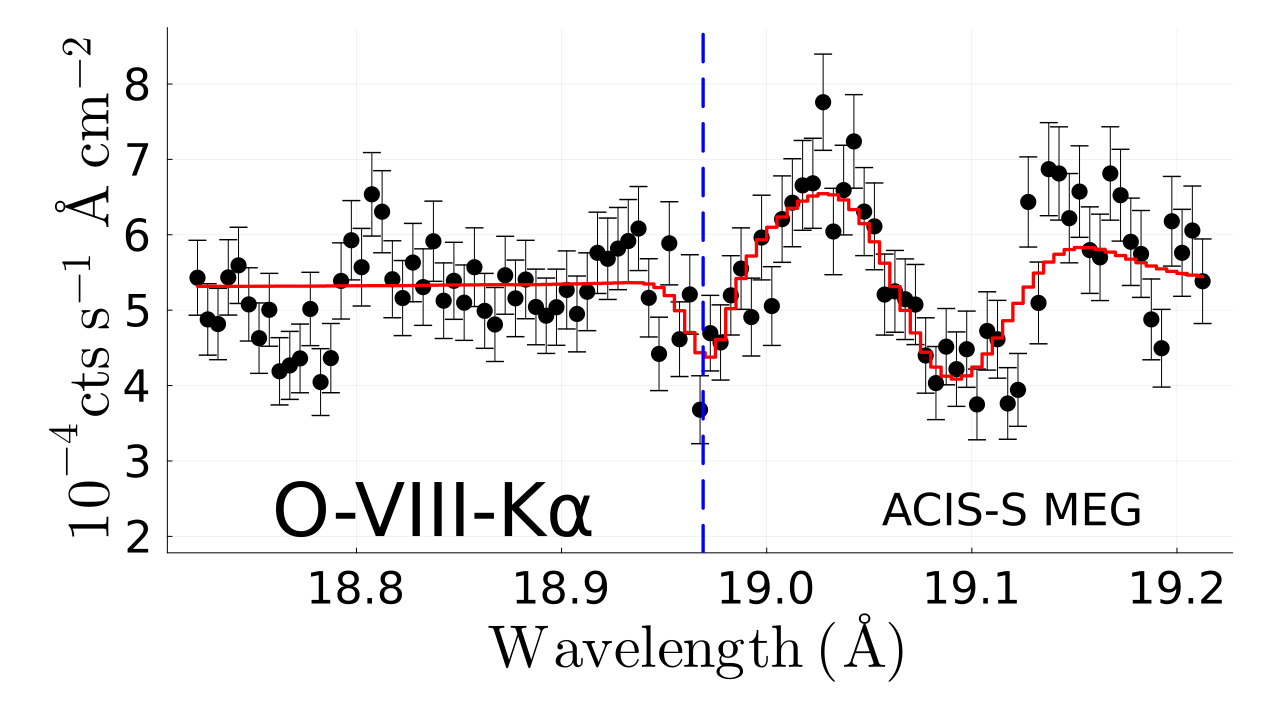}  &
            \includegraphics[width=0.33\textwidth]{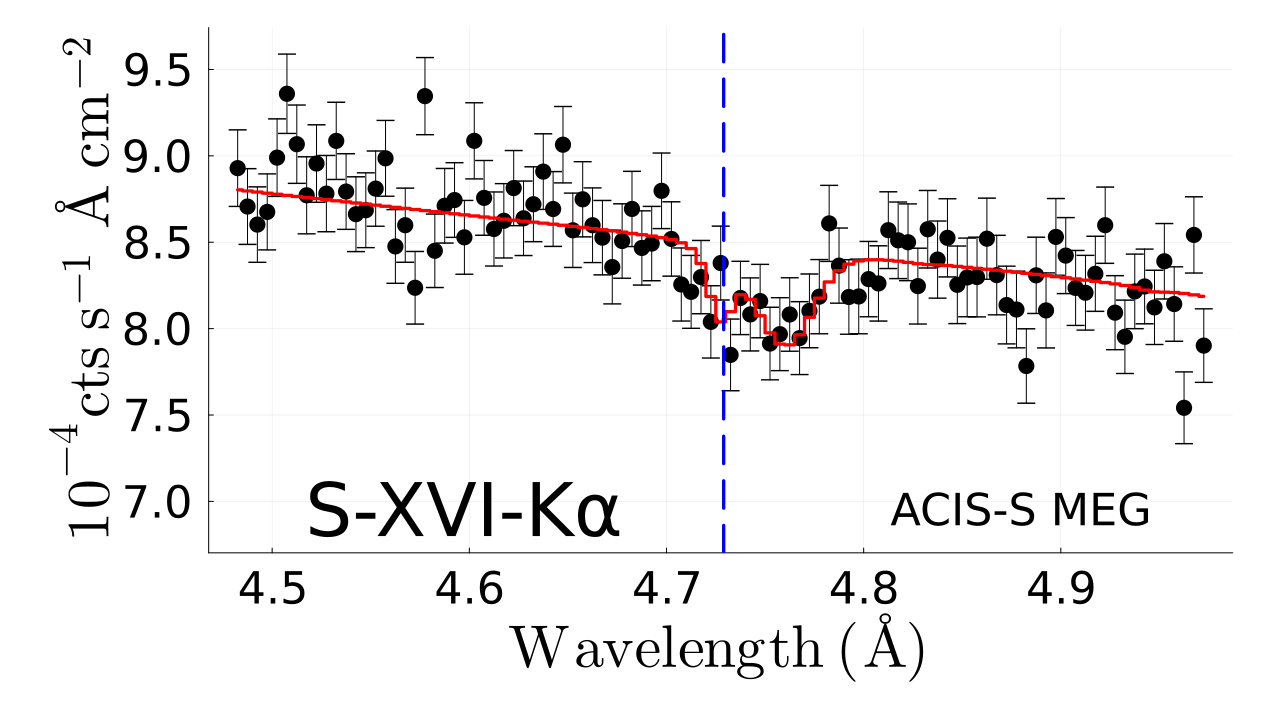}   &
            \includegraphics[width=0.33\textwidth]{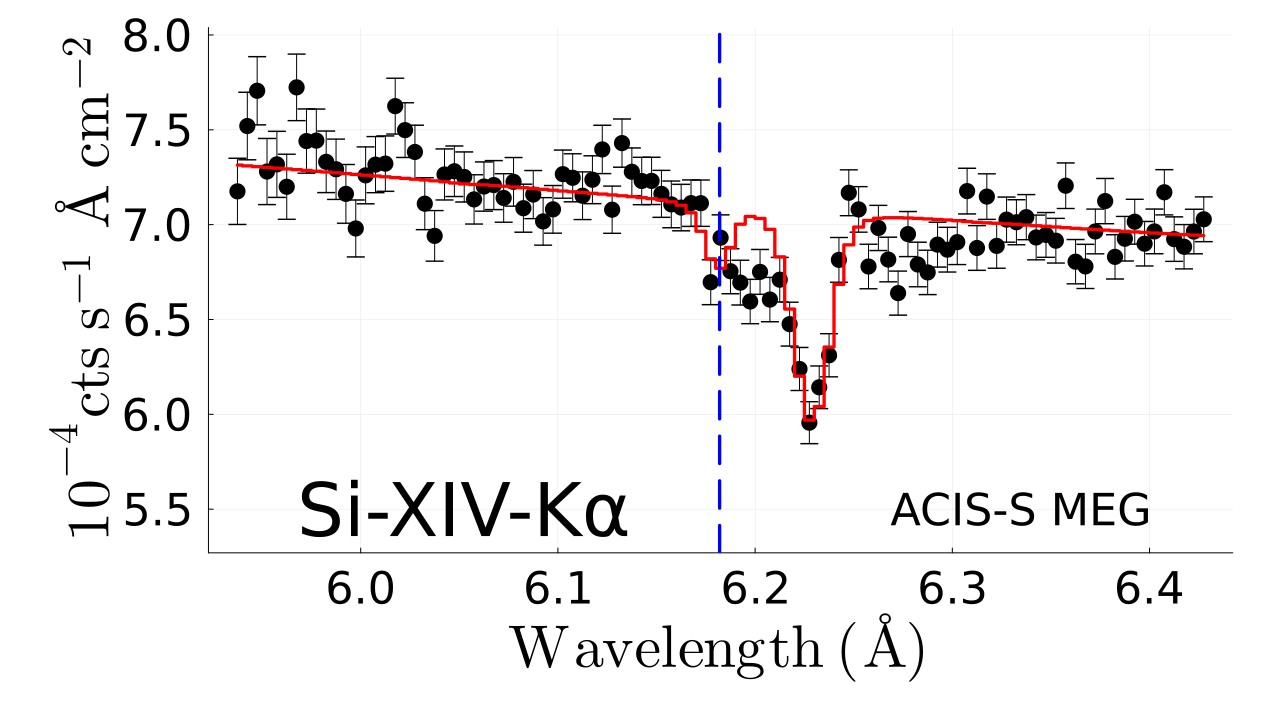}  \\   
    \end{tabular}
    \caption{We present the plots of absorption lines for different ionic transitions in \MEG\ data modeled with \PHASE. The data points are represented by black markers, the best-fit model is indicated by a red line, and the position of the theoretical rest wavelength is denoted by a vertical blue dashed line.}
    \label{fig:linesMEG}
    \end{figure*}

\begin{figure*}
        \centering
        \begin{tabular}{ccc}
            \includegraphics[width=0.33\textwidth]{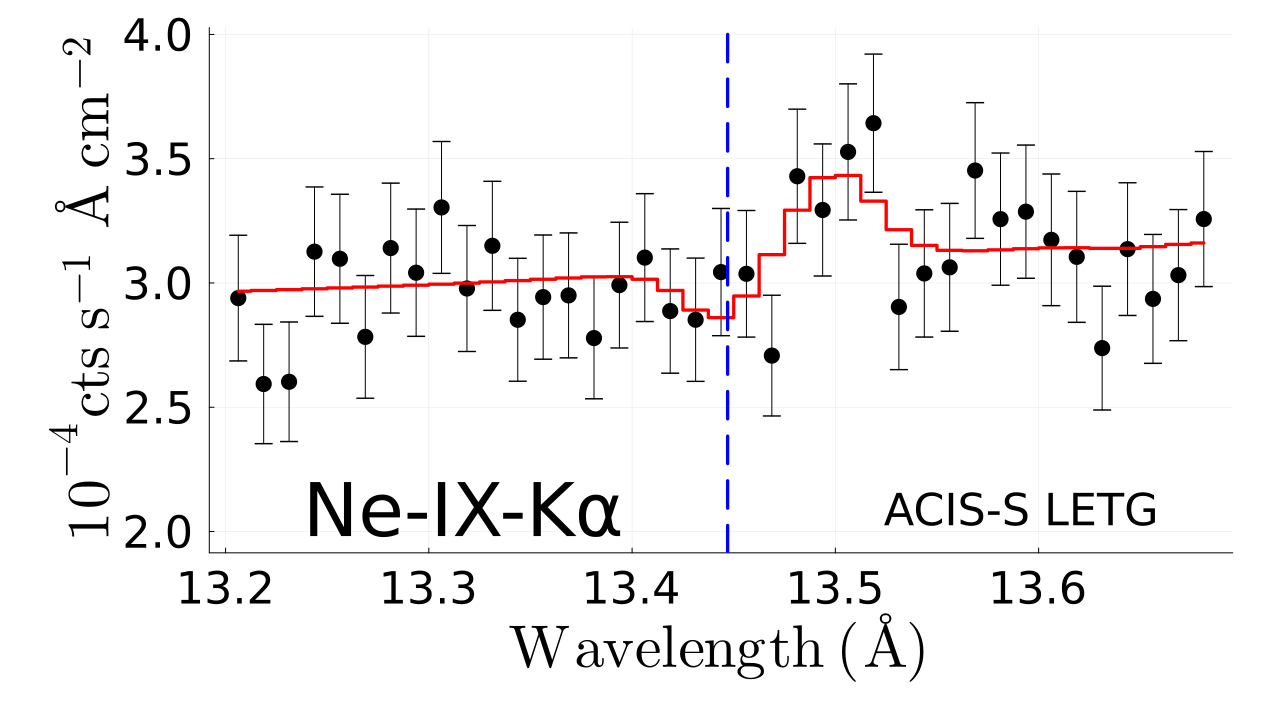}   &
            \includegraphics[width=0.33\textwidth]{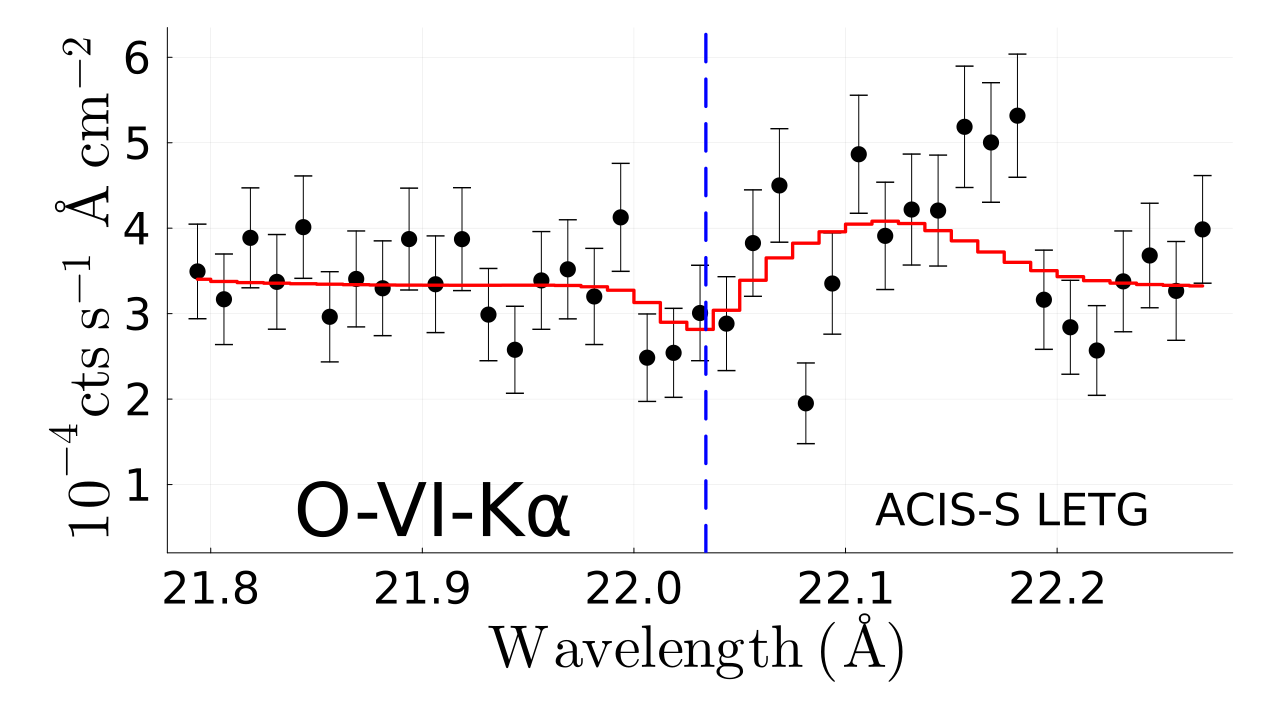}    &
            \includegraphics[width=0.33\textwidth]{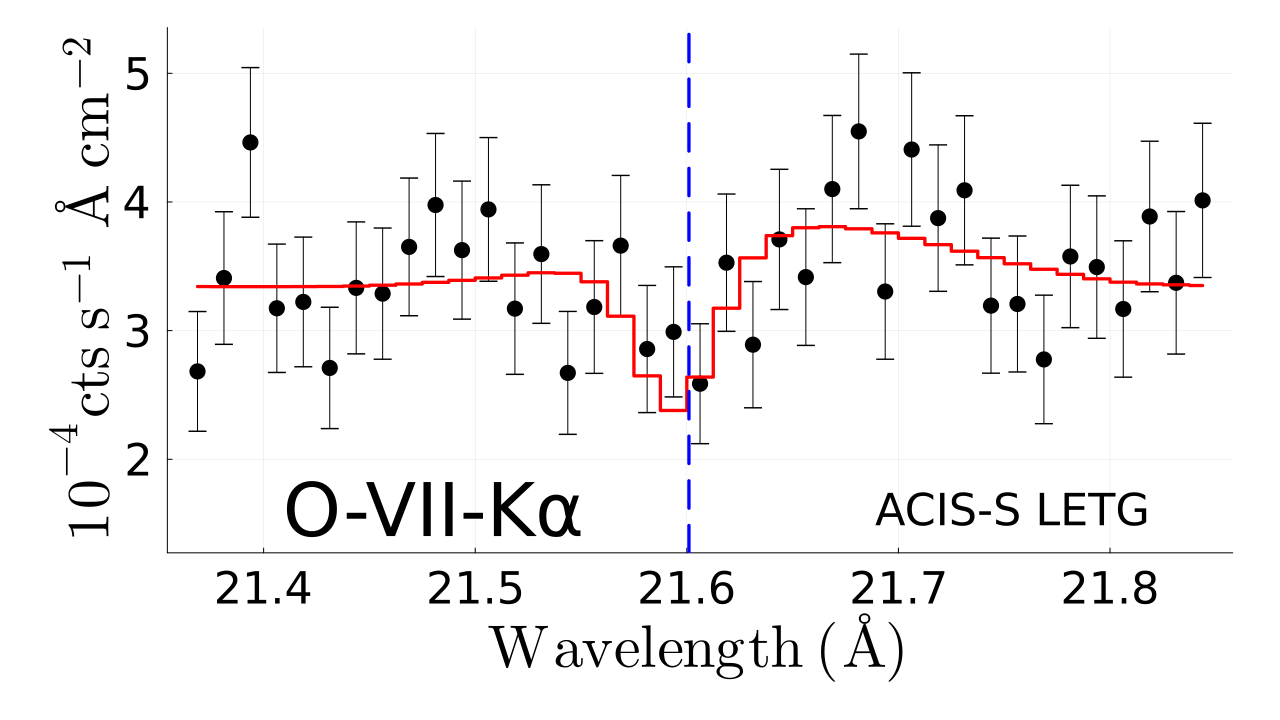}   \\
            \includegraphics[width=0.33\textwidth]{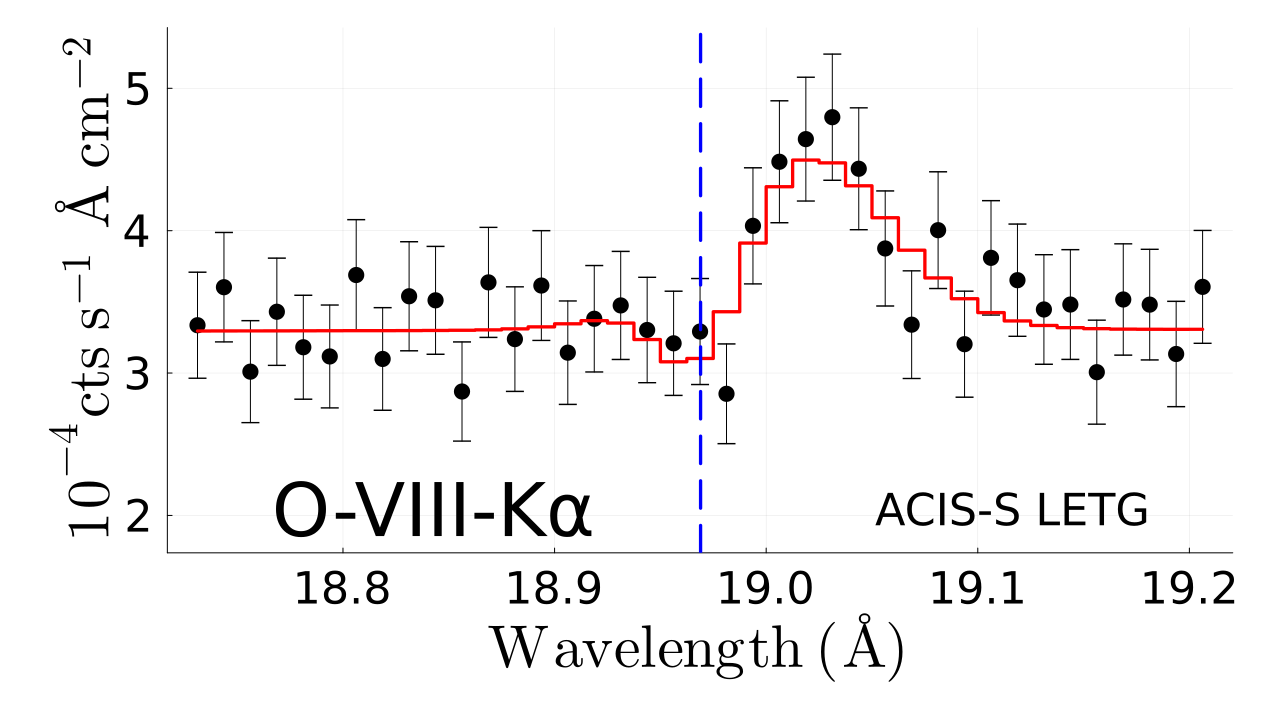}  &
            \includegraphics[width=0.33\textwidth]{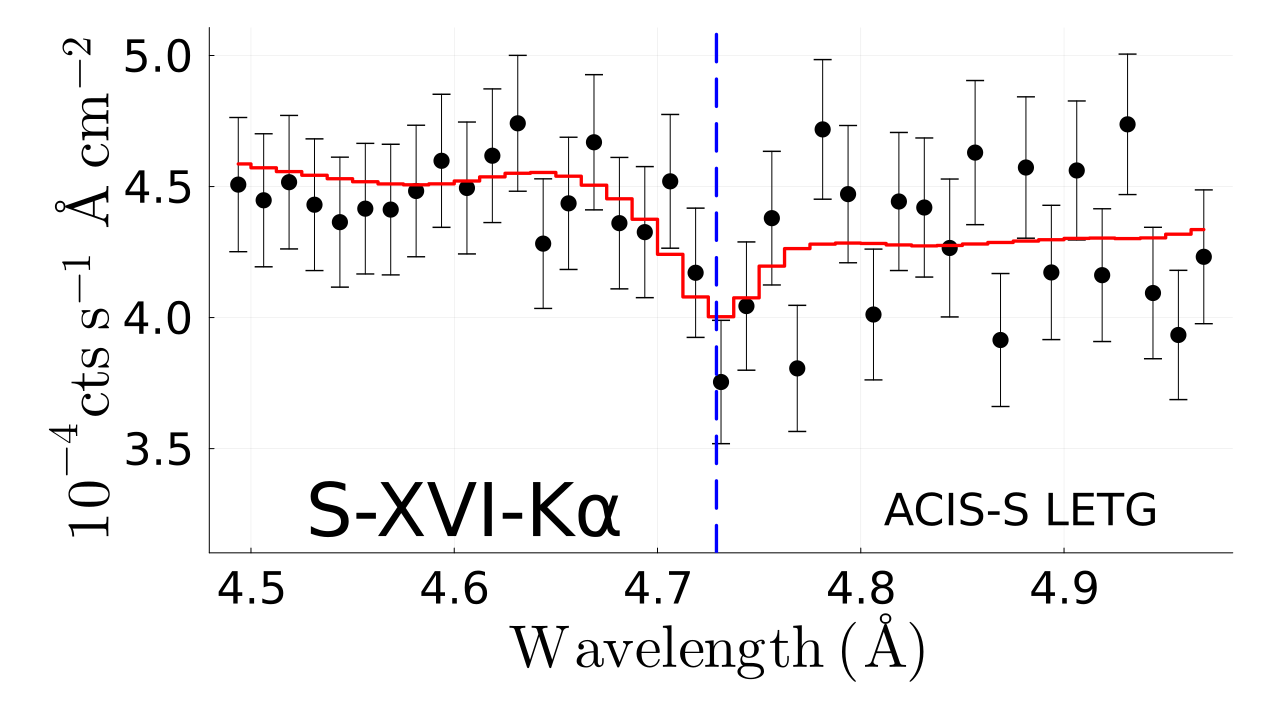}   &
            \includegraphics[width=0.33\textwidth]{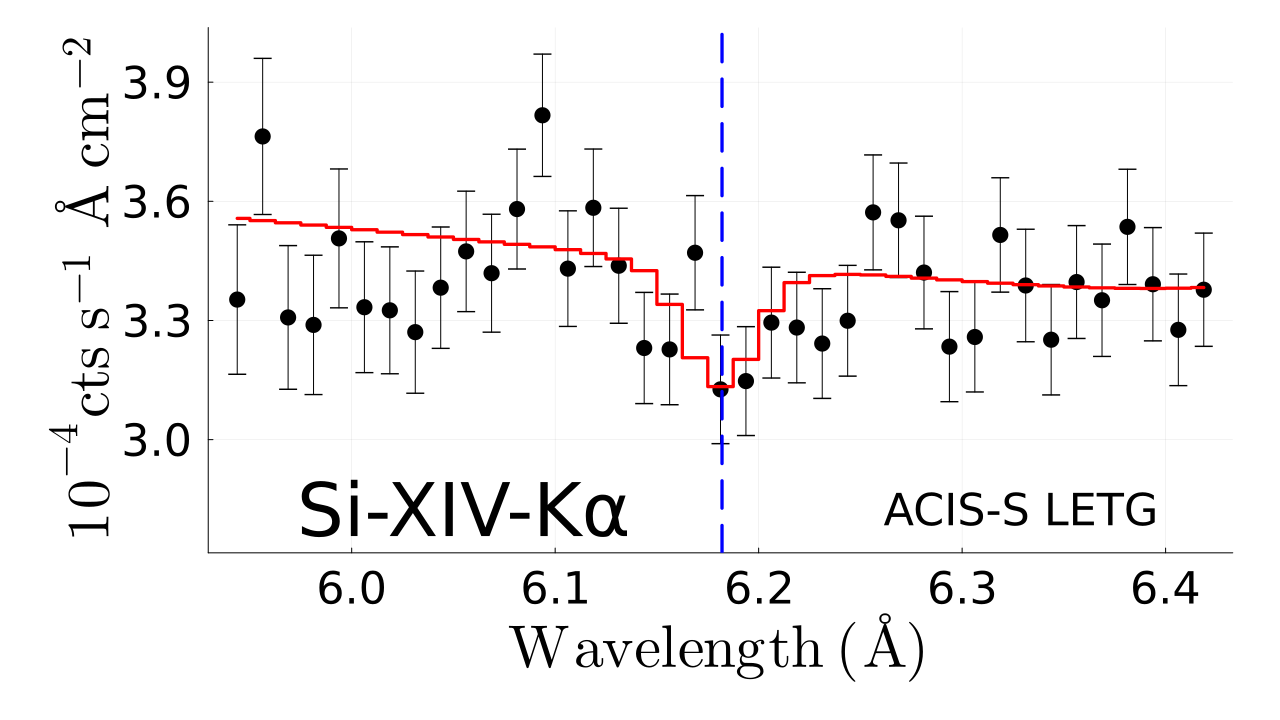}  \\
    \end{tabular}
    \caption{We present the plots of absorption lines for different ionic transitions in \LETG\ data modeled with \PHASE. The data points are represented by black markers, the best-fit model is indicated by a red line, and the position of the theoretical rest wavelength is denoted by a vertical blue dashed line.}
    \label{fig:linesLETG}
    \end{figure*}

        Figures~\ref{fig:linesMEG} and \ref{fig:linesLETG} show the lines fitted with \PHASE. The best model fitting \MEG\ and \LETG\ data comprises three different gas components. Each component models different absorption lines in the rest frame of the Milky Way. According to our models, some ionic transitions span through two different temperature components, while others are produced in a single gas phase.

        In Table~\ref{tab:results}, we present the physical parameters of the three components modeled in \MEG\ and \LETG\ datasets. We display the position of the absorption line, its Equivalent Width (EW) and its Ionic Column Density (N$_\mathrm{ion}$). Next to \SiXIV\ \Ka\ and \SXVI\ \Ka\ parameters in the \hot\ component, we also included in columns (12) and (14) the EW and the ionic column density of the lines as reported in LDI-2023a.


    \subsection{ACIS-S HETG-MEG} \label{sec:resultsHETG} 
       
        The first component modeling \MEG\ data is a \warmhot\ component at \logT\ = \TempMEGWARMHOT\ and column density \logNH\ = \NHMEGWARMHOT\ (hereafter \MEGWARMHOT). The absorption lines modeled with this component are \NVI\ \Kb, \NVII\ \Ka, \NeIX\ \Ka, \OVII\ \Ka, \OVII\ \Kb, \OVII\ \Kg, and \OVIII\ \Ka. This component improves the fit by a \delchi\ of \delchiMEGWARMHOT\ for two free pareameters. 

        The second component is a \hot\ component at \logT\ = \TempMEGSUPERHOT\ and column density \logNH\ = \NHMEGSUPERHOT\ (hereafter \MEGSUPERHOT). The absorption lines modeled with this component are \OVIII\ \Ka, \SiXIV\ \Ka, and \SXVI\ \Ka. This component improves the fit by a \delchi\ of \delchiMEGHOT\ for two additional free parameters
    
        The third component comprising the model fitting \MEG\ data is a \warm\ component at \logT\ = \TempMEGWARM\ and column density \logNH\ = \NHMEGWARM\ (hereafter \MEGWARM). Only \NVI\ \Kb\ absorption line is modeled with this component improving the fit marginally by a \delchi\ of \delchiMEGWARM\ for two additional free parameters, see Table~\ref{tab:results_stats}.
    
        \NVI\ \Kb\ contributes on the \warm\ and \warmhot\ component, while \OVIII\ \Ka\ is contributing in the \warmhot\ and \hot\ components. On the other hand, \NVII\ \Ka, \NeIX\ \Ka, \OVII\ \Ka, \OVII\ \Kb, \OVII\ \Kg\ contribute exclusively to the \warmhot\ component. \SiXIV\ and \SXVI\ contribute exclusively to the \hot\ component.
                    

    \subsection{ACIS-S LETG} \label{sec:resultsLETG} 

        The best model fitting \LETG\ data comprises three distinct gas components. The first component (hereafter \LETGWARMHOT) represents a \warmhot\ component with a temperature of \logT\ = \TempLETGWARMHOT\ and a column density of \logNH = \NHLETGWARMHOT. This component models \NeIX\ \Ka, \OVII\ \Ka, and \OVIII\ \Ka\ absorption lines. This component improves the fit only by a \delchi\ of \delchiLETGWARMHOT\ for two additional free parameters.

        The second and hotter component modeling \LETG\ data is a gas component with a temperature \logT\ = \TempLETGSUPERHOT\ and a column density of \logNH = \NHLETGSUPERHOT. This component is referred here as \LETGSUPERHOT. The absorption lines modeled with this component are \OVIII\ \Ka, \SXVI\ \Ka, and \SiXIV\ \Ka. This component improves the fit by a \delchi\ of \delchiLETGHOT\ for two additional free parameters.
        
        Finally, the third component fitting this dataset is characterized by a temperature of \logT\ = \TempLETGWARM\ and a column density of \logNH = \NHLETGWARM, (hereafter \LETGWARM). The absorption lines modeled in this component are \OVI\ \Ka\ and \OVII\ \Ka. This component improves the fit by a \delchi\ of \delchiLETGWARM\ for two additional free parameters, see Table~\ref{tab:results_stats}.
    
        In our results for the \LETG\ data, we find \OVI\ \Ka\ contributing exclusively to \LETGWARM. \OVII\ is present in the \LETGWARM\ and \LETGWARMHOT\ components. \NeIX\ is found exclusively in the \LETGWARMHOT\ component. \OVIII\ contributes in both, \LETGWARMHOT\ and \LETGSUPERHOT, while \SiXIV\ and \SXVI\ are exclusively found in \LETGSUPERHOT.
        

    \subsection{Metallicity of the CGM}

\begin{figure}
        \centering
        \renewcommand{\arraystretch}{2}\addtolength{\tabcolsep}{2pt}
        \begin{tabular}{c}
            \includegraphics[width=0.45\textwidth]{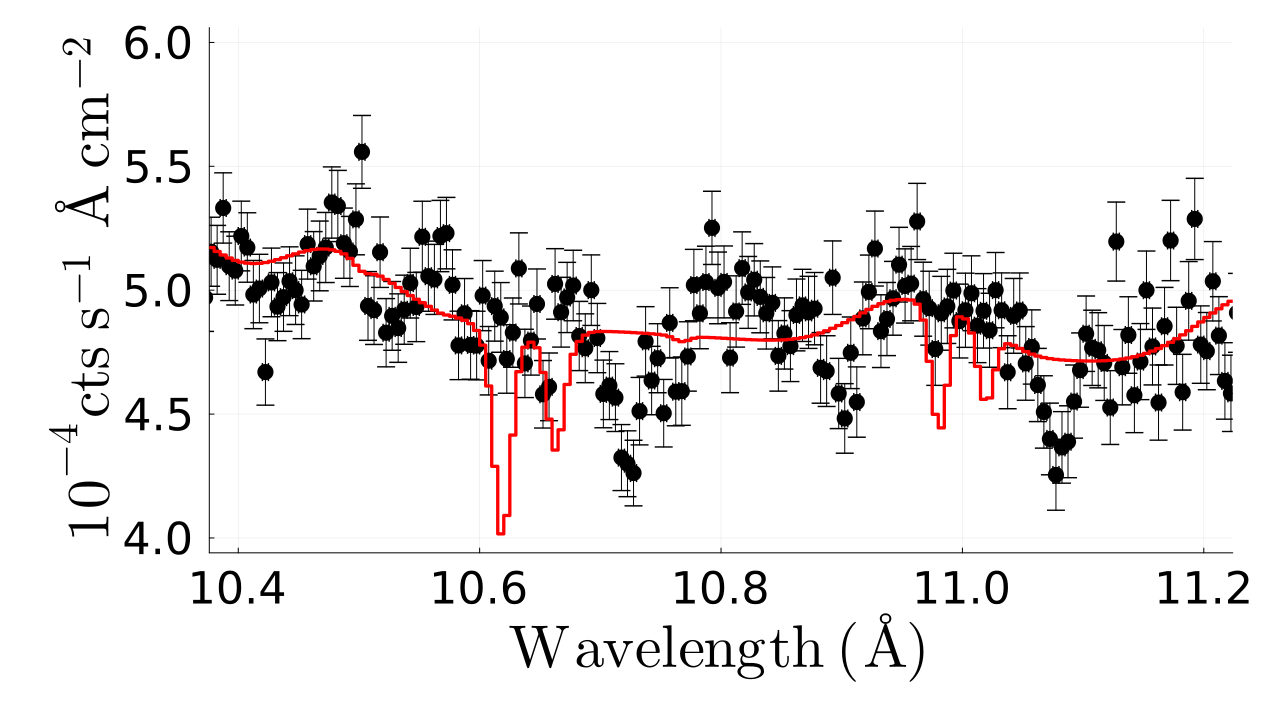}    \\
            \includegraphics[width=0.45\textwidth]{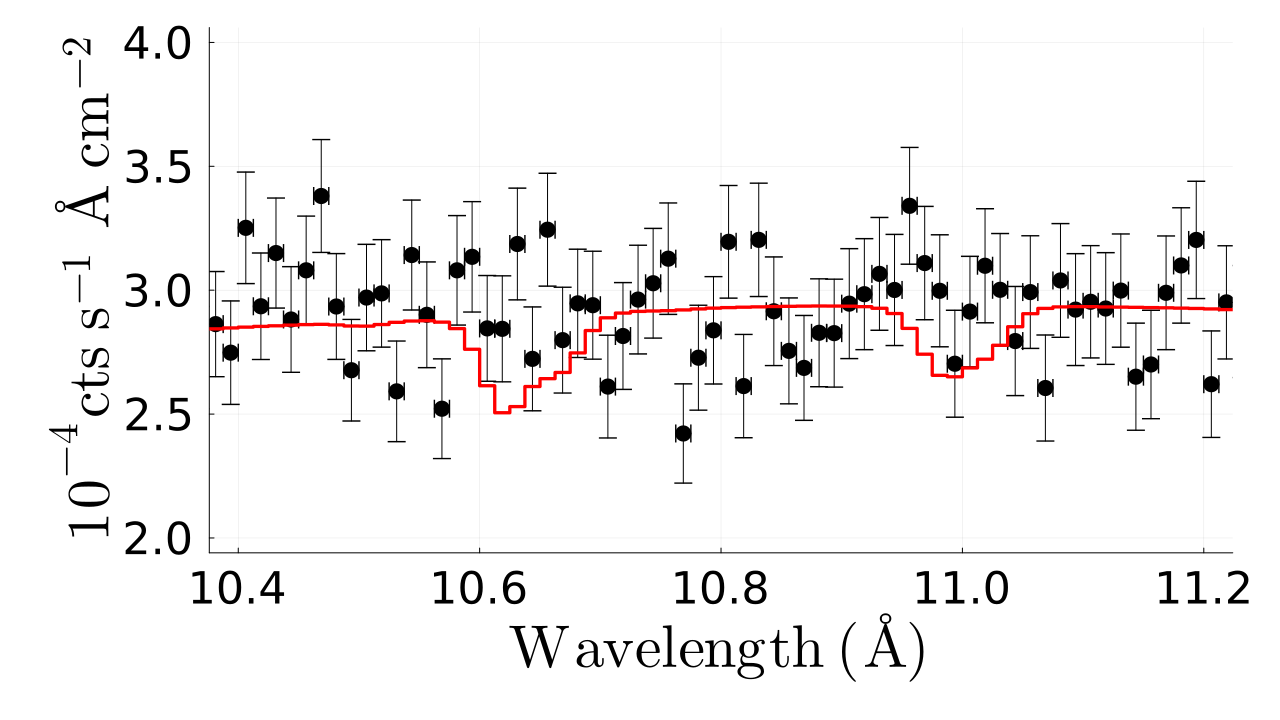}   \\
        \end{tabular}
        \caption{These plots present the model in red overpredicting \Fe\ L-shell absorption lines in \MEG\ (top) and \LETG\ (bottom) data when \Fe\ abundance is set to be that of Oxygen.}
        \label{fig:Fe}
    \end{figure}

        In \MEG\ data, we could not model \MEGSUPERHOT\ with solar \Fe\ abundance since it grossly overpredicts the \Fe\ absorption lines. In Figure~\ref{fig:Fe}, we present the spectrum of \MEG\ data in the range 10.4 - 11.2 \AA. In this Figure, we show how the model overpredicts the absorption lines of \Fe\ when its abundance is that of Oxygen. To avoid this, the model must consider an abundance of \Fe\ at least four times less than that of Oxygen. We find the abundance of other elements correspond to solar mixture.

        In \LETG\ data, the model also overpredicts the \Fe\ L-shell lines in the \hot\ component. Figure~\ref{fig:Fe} shows how the model overpredicts \Fe\ when its abundance is that of Oxygen. In this case, the abundance of \Fe\ in the \hot\ component needs to be at least six times less than that of Oxygen to properly model the data. 

        We also noticed that for \LETG\, the code prefers $\sim$ 4 times more \Sulfur/\Oxygen\ than the solar value. In this same dataset, the code prefers $\sim$ 1.5 times solar \Si/\Oxygen. Nonetheless, the data is not very sensitive to the abundance of these two elements.
 

    \begin{table}
        \centering
        \caption{\delchi\ improved in \MEG\ and \LETG\ when including a component modeling a different gas phase with two free parameters.}
    \begin{tabular}{llll}
    \hline
    Component & \logT\ & \logNH & \delchi \\
    \hline
    \MEGWARM & \TempMEGWARM & \NHMEGWARM & \delchiMEGWARM \\
    \MEGWARMHOT & \TempMEGWARMHOT & \NHMEGWARMHOT & \delchiMEGWARMHOT \\
    \MEGSUPERHOT & \TempMEGSUPERHOT & \NHMEGSUPERHOT & \delchiMEGHOT \\
    \LETGWARM & \TempLETGWARM & \NHLETGWARM & \delchiLETGWARM \\
    \LETGWARMHOT & \TempLETGWARMHOT & \NHLETGWARMHOT & \delchiLETGWARMHOT \\
    \LETGSUPERHOT & \TempLETGSUPERHOT & \NHLETGSUPERHOT & \delchiLETGHOT \\
    \hline
    \end{tabular}
    \label{tab:results_stats}
    \end{table}


    \begin{table*}
        \centering
        \caption{Identified ionic absorption lines probing the super-virial hot component at z=0.}
        \resizebox{\textwidth}{!}{%
        {\begin{tabular}{llc|ccc|ccc|ccccc}
        \toprule
        Ion
                        &   Tran
                        &   RestWav (\AA)
                        &   Wav (\AA)
                        &   EW (m\AA)
                        &   N $_\mathrm{ion}$ (cm$^{-2}$)
                        &  Wav (\AA)
                        &   EW (m\AA)
                        &   N $_\mathrm{ion}$ (cm$^{-2}$)
                        &   Wav (\AA)
                        &   EW (m\AA)
                        &   EW (m\AA)
                        &   N $_\mathrm{ion}$ (cm$^{-2}$)
                        &   N $_\mathrm{ion}$ (cm$^{-2}$)\\
                        (1) & (2) & (3)& (4)& (5)& (6)& (7)& (8)& (9)& (10)& (11)& (12)& (13) &(14)\\
                        \toprule
            \\
            \\
        \multicolumn{3}{c}{{}}
                        &   \multicolumn{3}{c}{{\textbf{\MEGWARM}}}
                        &   \multicolumn{3}{c}{{\textbf{\MEGWARMHOT}}}
                        &   \multicolumn{5}{c}{{\textbf{\MEGSUPERHOT}}} \\
                            \multicolumn{3}{c}{{}}
                        &   \multicolumn{3}{c}{{\fontfamily{qcr}\selectfont \logT\ = \TempMEGWARM}}
                        &   \multicolumn{3}{c}{{\fontfamily{qcr}\selectfont \logT\ = \TempMEGWARMHOT}}
                        &   \multicolumn{5}{c}{{\fontfamily{qcr}\selectfont \logT\ = \TempMEGSUPERHOT}}  \\
                            \multicolumn{3}{c}{{}}
                        &   \multicolumn{3}{c}{{\fontfamily{qcr}\selectfont \logNH\ = \NHMEGWARM}}
                        &   \multicolumn{3}{c}{{\fontfamily{qcr}\selectfont \logNH\ = \NHMEGWARMHOT}}
                        &   \multicolumn{5}{c}{{\fontfamily{qcr}\selectfont \logNH\ = \NHMEGSUPERHOT}} \\ \\
        \NVI &\Kb &\PosNVIKb &\PosNVIKbMEGWARM &\EWNVIKbMEGWARM &\NHNVIMEGWARM &\PosNVIKbMEGWARMHOT &\EWNVIKbMEGWARMHOT &\NHNVIMEGWARMHOT & & & & &\\
        \NVII &\Ka &\PosNVIIKa & & & &\PosNVIIKaMEGWARMHOT &\EWNVIIKaMEGWARMHOT &\NHNVIIMEGWARMHOT & & & & &\\
        \NeIX &\Ka &\PosNeIXKa & & & &\PosNeIXKaMEGWARMHOT &\EWNeIXKaMEGWARMHOT &\NHNeIXMEGWARMHOT & & & & &\\
        \OVII &\Ka &\PosOVIIKa & & & &\PosOVIIKaMEGWARMHOT &\EWOVIIKaMEGWARMHOT &\NHOVIIMEGWARMHOT & & & & &\\
        \OVII &\Kb &\PosOVIIKb & & & &\PosOVIIKbMEGWARMHOT &\EWOVIIKbMEGWARMHOT &\NHOVIIMEGWARMHOT & & & & &\\
        \OVII &\Kg &\PosOVIIKg & & & &\PosOVIIKgMEGWARMHOT &\EWOVIIKgMEGWARMHOT &\NHOVIIMEGWARMHOT & & & & &\\
        \OVIII &\Ka &\PosOVIIIKa & & & &\PosOVIIIKaMEGWARMHOT &\EWOVIIIKaMEGWARMHOT &\NHOVIIIMEGWARMHOT &\PosOVIIIKaMEGHOT &\EWOVIIIKaMEGHOT & &\NHOVIIIMEGHOT &\\
        \SXVI &\Ka &\PosSXVIKa & & & & & & &\PosSXVIKaMEGHOT &\EWSXVIKaMEGHOT & (\EWSXVIKaMEGLaraDI) &\NHSXVIMEGHOT & (\NHSXVIKaMEGLaraDI) \\
        \SiXIV &\Ka &\PosSiXIVKa & & & & & & &\PosSiXIVKaMEGHOT &\EWSiXIVKaMEGHOT & (\EWSiXIVKaMEGLaraDI) &\NHSiXIVMEGHOT &(\NHSiXIVKaMEGLaraDI)\\
            \\
            \\
            \multicolumn{3}{c}{{}}
                        &   \multicolumn{3}{c}{{\textbf{\LETGWARM}}}
                        &   \multicolumn{3}{c}{{\textbf{\LETGWARMHOT}}}
                        &   \multicolumn{5}{c}{{\textbf{\LETGSUPERHOT}}} \\
                            \multicolumn{3}{c}{{}}
                        &   \multicolumn{3}{c}{{\fontfamily{qcr}\selectfont \logT\ = \TempLETGWARM}}
                        &   \multicolumn{3}{c}{{\fontfamily{qcr}\selectfont \logT\ = \TempLETGWARMHOT}}
                        &   \multicolumn{5}{c}{{\fontfamily{qcr}\selectfont \logT\ = \TempLETGSUPERHOT}}  \\
                            \multicolumn{3}{c}{{}}
                        &   \multicolumn{3}{c}{{\fontfamily{qcr}\selectfont \logNH\ = \NHLETGWARM}}
                        &   \multicolumn{3}{c}{{\fontfamily{qcr}\selectfont \logNH\ = \NHLETGWARMHOT}}
                        &   \multicolumn{5}{c}{{\fontfamily{qcr}\selectfont \logNH\ = \NHLETGSUPERHOT}} \\ \\
            \NeIX &\Ka &\PosNeIXKa & & & &\PosNeIXKaLETGWARMHOT &\EWNeIXKaLETGWARMHOT &\NHNeIXLETGWARMHOT & & & & &\\
            \OVI &\Ka &\PosOVIKa &\PosOVIKaLETGWARM &\EWOVIKaLETGWARM &\NHOVILETGWARM & & & & & & & &\\
            \OVII &\Ka &\PosOVIIKa &\PosOVIIKaLETGWARM &\EWOVIIKaLETGWARM &\NHOVIILETGWARM &\PosOVIIKaLETGWARMHOT &\EWOVIIKaLETGWARMHOT &\NHOVIILETGWARMHOT & & & & &\\
            \OVIII &\Ka &\PosOVIIIKa & & & &\PosOVIIIKaLETGWARMHOT &\EWOVIIIKaLETGWARMHOT &\NHOVIIILETGWARMHOT &\PosOVIIIKaLETGHOT &\EWOVIIIKaLETGHOT & &\NHOVIIILETGHOT &\\
            \SXVI &\Ka &\PosSXVIKa & & & & & & &\PosSXVIKaLETGHOT &\EWSXVIKaLETGHOT &(\EWSXVIKaLETGLaraDI) &\NHSXVILETGHOT &(\NHSXVIKaLETGLaraDI) \\
            \SiXIV &\Ka &\PosSiXIVKa & & & & & & &\PosSiXIVKaLETGHOT &\EWSiXIVKaLETGHOT &(\EWSiXIVKaLETGLaraDI) &\NHSiXIVLETGHOT &(\NHSiXIVKaLETGLaraDI)\\
        \end{tabular}}}
        \label{tab:results}
    \end{table*}


    \begin{figure}
        \centering
        \renewcommand{\arraystretch}{2}\addtolength{\tabcolsep}{2pt}
        \begin{tabular}{c}
            \includegraphics[width=0.45\textwidth]{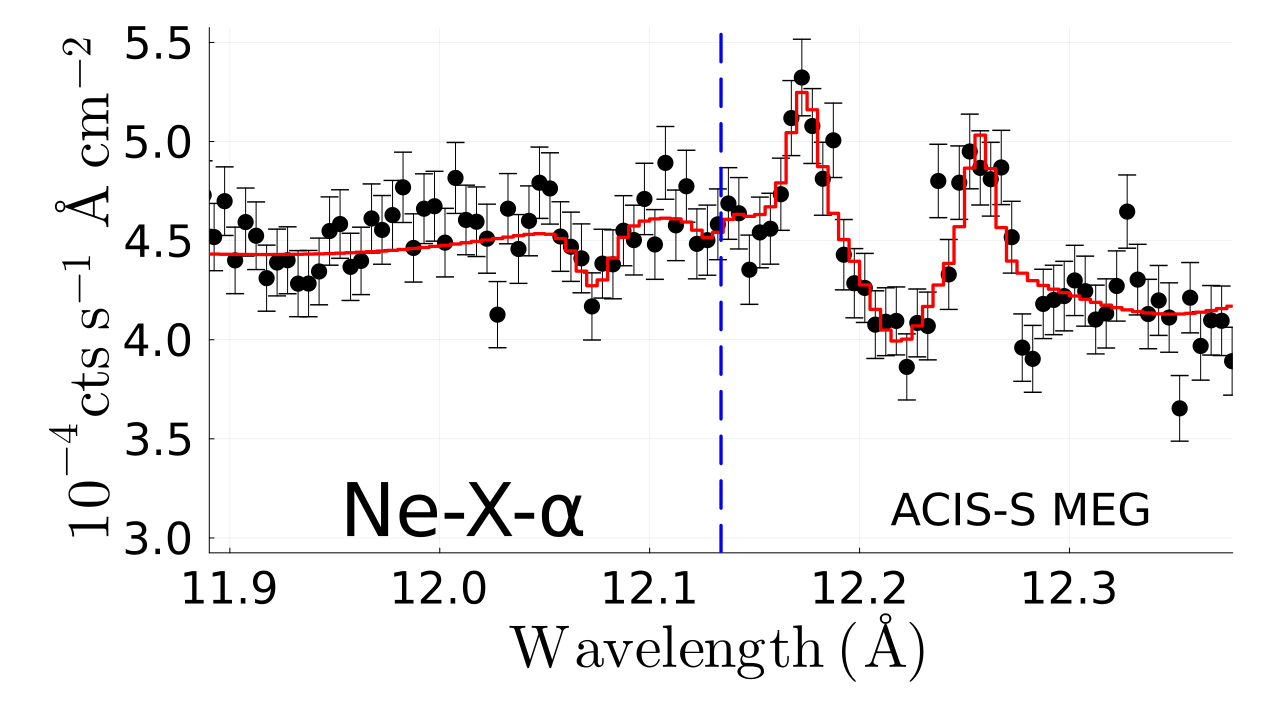}    \\
            \includegraphics[width=0.45\textwidth]{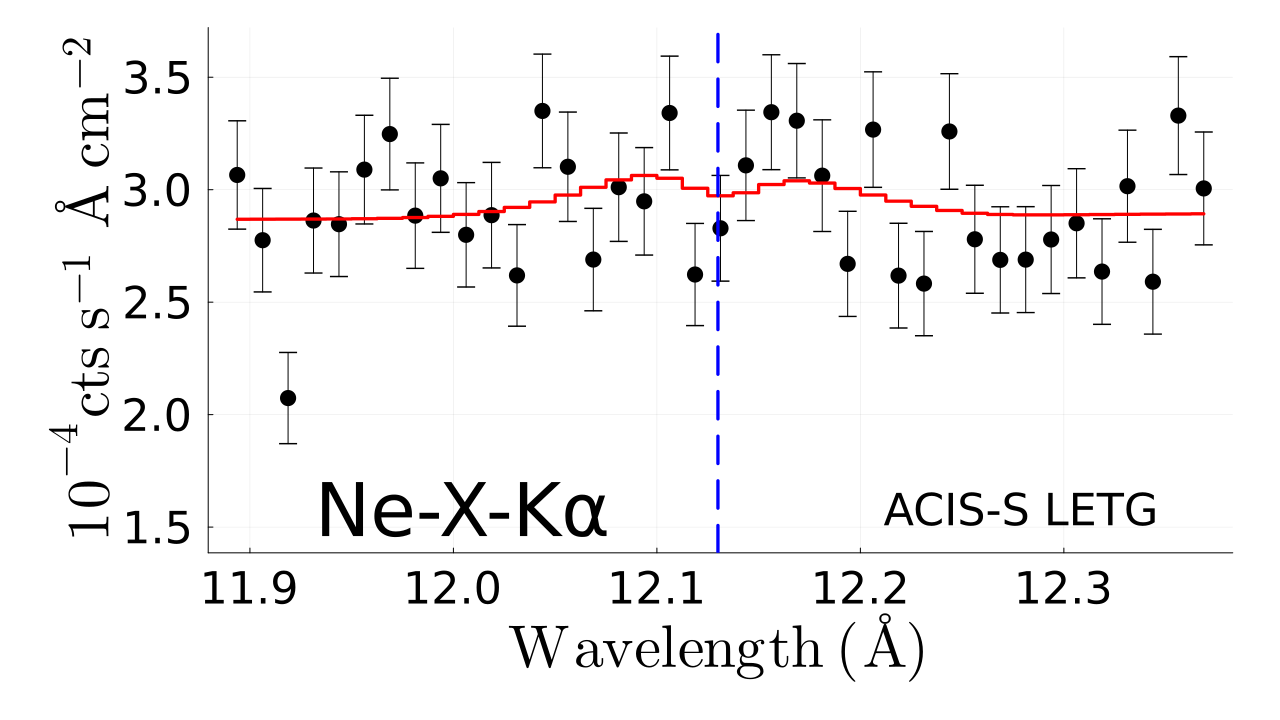}   \\
        \end{tabular}
        \caption{Here we present a spectral window of our \MEG\ (top) and \LETG\ (bottom) data modeled. We show in vertical blue line the position of  \Ne\ \Ka\ res-wavelength}
        \label{fig:Ne}
    \end{figure}

%
%

    \section{Discussion \label{sec:discussion}}

        We have found that the Milky Way's CGM comprises at least three phases, a \warm\ (\logT\ $\sim$ 5.5), \warmhot\ (\logT\ $\sim$ 6), and \hot\ (\logT\ $\sim$ 7.5) component, confirming previous results pointing to the third phase with high temperature (Das et al. 2019 and 2021; LDI-2023a). This \hot\ phase has a column density at least an order of magnitude higher than the \warmhot\ and \warm\ components. This result is consistent in \MEG\ and \LETG\ data and previous detections of the \hot\ phase (Das et al. 2019 and 2021; LDI-2023a).


        \subsection{Physical state of the gas in the \hot\ component}
            
            It is interesting to note that we do not see \NeX\ \Ka\ in \MEG\ or \LETG. This is consistent with the high temperature of the \hot\ component. \NeX\ peaks around \logT\ $\sim$ 6.7, and the fraction of \NeX\ decreases sharlply for higher temperatures. At \logT\ $\sim$ 7.5, \Ne\ is almost completely ionized. Therefore, our data is consistent with not being able to detect \NeX\ in \MEG\ or \LETG, and this is a nice confirmation that the temperature of the gas reaches \logT\ $\sim$ 7.5. The \NeX\ line should be very weak due to the plasma being very hot. Therefore, it is not detectable with the S/N of our data. In Figure~\ref{fig:Ne} we show our model fitting the data at the position of the rest-wavelength of \NeX. We see no prominent absoprtion line of this ion. \citet{Das2019} and \citet{Das2021} reported the detection of this line at the same temperature, but their model required super-solar \Ne/\Oxygen\ abundance. If the plasma had been a factor of two or three lower temperature, say at \logT\ $\sim$ 7, the \NeX\ line should be very prominent and detectable in our spectra. This result excludes the presence of high quantities of gas in the CGM with temperatures \logT\ over $\sim$ 6.5 and below $\sim$ 7.5 in the sightlines probed in the stacked spectra. 


        \subsection{Comparison between detections of LDI-2023a and this work}

            As mentioned in \S\ \ref{sec:intro}, LDI-2023a clearly detected \SiXIV\ \Ka\ and \SXVI\ \Ka\ in the same stacked spectra presented here. In Table~\ref{tab:results}, we show the EW and N$_\mathrm{ion}$ predicted by our model with \PHASE, contrasted with those measured empirically by LDI-2023a over \MEG\ and \LETG\ with Gaussians. The striking similarity in these values shows that detecting these elements is a robust result in these studies. Therefore, the results presented here fully confirm, with a self-consistent ionization model, the presence of a \hot\, super-virial gas component in the Milky Way's CGM.


        \subsection{Statistical significance of each component}

            In Table~\ref{tab:results_stats}, we show how much each component changes the $\chi^2$ statistic for two free parameters (namely, the temperature and the column density of the gas). For \MEG\ data, we find that the gas components weighting more in the statistic are \MEGWARMHOT\ with \delchi\ = \delchiMEGWARMHOT\ and \MEGHOT\ with \delchi\ = \delchiMEGHOT. The fact that these phases are statistically significant is not a surprise, given the number of absorption lines fit with \MEGWARMHOT\ component and given the high column density of \MEGHOT\ component. On the other hand, the \MEGWARM\ only changes the statistic by a \delchi\ = \delchiMEGWARM. We note, however, that there is vast evidence of the presence of this component in UV data (e.g., \citealp{Tumlinson2017} and references therein). 

            In \LETG\, we find that the \LETGWARM\ and \LETGHOT\ weight similarly with \delchi\ = \delchiLETGWARM\ and \delchi\ = \delchiLETGHOT\ respectively, making both components statistically significant. On the other hand, \LETGWARMHOT\ component only weights \delchi\ = \delchiLETGWARMHOT\ in our model, making it barely significant. This can be attributed to the low S/N ratio of these data, given that the column density of this component found in \MEG\ and \LETG\ are similar. We notice, however, that the \OVII\ \Ka\ line is clearly detected in the data but with less significance ($\sigma$ $\approx$ 2). 
            

        \subsection{Comparison between \MEG\ and \LETG\ models}
        
            The results from \MEG\ and \LETG\ are consistent. In the \warmhot\ component, the model over both datasets has similar column densities (within the errors) and temperatures. These parameters are strikingly similar to what has been reported in individual lines of sight (e.g., \citealp{Gupta2012}). Thus, our results bring more evidence to the homogeneity of this component. This is in contrast to the results found in emission, where a large dispersion up by an order of magnitude has been reported in the temperature (e.g., \citealp{Henley2010};  \citealp{Bluem2022}; \citealp{Gupta2023} and references therein; \citealp{Bhattacharyya2023}). 
 
            The same happens with the temperature of the \hot\ component, which is consistent within the errors between the two datasets. However, there is a factor of $\sim$ 3 in the column densities between the data of the two instruments. We note that only one source in \LETG\ is not present in \MEG\ stacked spectrum. Nevertheless, there are 38 more sources in \MEG\ than in \LETG. Therefore, this result, along with the consistency between our findings and those found in individual lines of sight (\citealp{Das2019}, \citealp{Das2021}), suggest that this component might also be homogeneous in temperature but with some dispersion in column density. 
       
            In the \warm\ component, there are some differences between \MEG\ and \LETG. These differences are marginal in temperature but become much larger in terms of column density. We find that this component is only constrained through one or two lines from one or two ions. Given this, and the low S/N ratio in the \LETG\ data, and the low significance of this component in the \MEG\ data, we consider that while we have enough evidence for the presence of this component (e.g., \citealp{Tumlinson2017}), we cannot constrain robustly its parameters. 


        \subsection{Atomic abundances in the \hot\ component \label{sec:AtomicAbundances}}

            We find additional evidence of inhomogeneities in the \hot\ component in addition to those in the column density reported above. In particular, the abundance of \SXVI\ is $\sim$ 3.5 times larger in \LETG\ than in \MEG. Additionally, we find that \Si/\Oxygen\ in \LETG\ is $\sim$ 2 times that in \MEG. The abundance of \Fe\ is strikingly below what is expected for solar mixture. As observed in Figure \ref{fig:Fe}, this is a strong observational constraint. In \LETG\, we required \Oxygen/\Fe\ {larger than solar} by at least a factor of 6, and in \MEG\ by at least a factor of 4. The sub-solar \Fe/\Oxygen\ was also found by \citet{Das2019} and \citet{Das2021}. Abundances for other atoms are consistent with solar mixture, but we notice that the sensitivity of both stacked spectra is very limited for many of the absorption transitions. 
        
            These results, along with other claims of the non-solar mixture in the \hot\ component (e.g., for \Ne/\Oxygen\ \citealp{Das2019}, \citealp{Das2021}, \citealp{Gupta2021}; for \Si/\Oxygen\ \citealp{Das2021}) suggest that it has inhomogeneous element abundance mixtures which are different than solar.


        \subsection{Dispersion Measure of the hot phase}

            In the present work, the column density of the hot phase as measured in \MEG\ data is

            \begin{equation}\label{eq:NH}
                N_H = 2\times10^{21} \mathrm{cm}^{-2} \left(\frac{A_O/A_H}{4.9 \times 10^{-4}}\right)^{-1} \left(\frac{Z/Z_\odot}{1}\right)^{-1},
            \end{equation}

            where the Oxygen abundance $(A_O/A_H)$ and metallicity $(Z)$ in the PHASE models discussed in this paper have been assumed to be solar ($A_O/A_H = 4.9 \times 10^{-4}$ \citet{Asplund2009} and $Z=1 Z_\odot$). The dispersion measure (DM), however, provides a constraint on the combination of these parameters.

            The DM of the CGM has been investigated using Fast Radio Bursts (FRBs) and Galactic Pulsars (e.g., \citealp{Prochaska2019, Platts2020, Das2021MNRAS, Cook2023}). \citealp{Platts2020} found {a DM for the Galactic halo ($DM_\mathrm{total}$) of} 
            
            \begin{equation}\label{eq:DMhaloplats}
                DM_\mathrm{total}=63^{+27}_{-21} \mathrm{\ pc \ cm}^{-3},
            \end{equation}

            {where the errors are $1\sigma$.}

            We calculated the contribution to the DM from the \warmhot\ component (DM$_\mathrm{virial}$) using equation (5) from \citet{Das2021MNRAS}. {By considering the \OVII\ column density ($N_{\text{OVII}} = 8.22^{+2.8}_{-2.97} \times 10^{15} \mathrm{cm}^{-2}$) and its ionic fraction (0.826) on the \MEGWARMHOT\ component, we obtained DM$_\mathrm{virial} = 7.9^{+2.8}_{-2.9} \mathrm{cm}^{-3} \mathrm{\ pc} .$ We then calculated the DM for the super-virial component using}

             \begin{equation}\label{eq:DMsupvir}
                DM_\mathrm{super-virial}  = DM_\mathrm{total}  - DM_\mathrm{virial}.
            \end{equation}

            {This results in DM$_\mathrm{super-virial}= 55.1^{+29.9}_{-23.7}$ pc cm$^{-3}$. Therefore, for satisfying Eq~\ref{eq:NH} and Eq~\ref{eq:DMsupvir} at the same time, $\left(\frac{A_o}{A_H}\right)\left(\frac{Z}{Z\odot}\right)=4.48^{+3.40}_{-1.57}\times10^{-3}$ is required.}

            These results implies that the observed hot gas is either enriched in Oxygen, Silicon and Sulfur (the elements driving the detection of the \hot\ component in this work), or has super-solar metallicity  or a combination of both. {For the best value of $DM_\mathrm{super-virial}=55.1$ pc cm$^{-3}$, and assuming solar abundance, the metallicity required would be $\sim$ 9 times solar, or over 6 solar assuming the 1$\sigma$ lower limit}. In \S~\ref{sec:AtomicAbundances} we already discussed the over solar mixture of these elements. This further suggests that the hot component finds its origin in galactic winds of enriched material. 
 

        \subsection{Contribution for \OII\ \Kb ?}

            Finally, we note that the rest wavelength of \OVI\ \Ka\ is \PosOVIKa\ \AA. This wavelength also corresponds to the rest-wavelength of \OII\ \Kb\ (\PosOIIKb\ \AA; \citealp{Mathur2017}). We are uncertain about the contribution of \OII\ \Kb\ to the \OVI\ \Ka\ absorption line. However, this possible contamination by \OII\ does not affect the detection of the \warm\ component since \NVI\ \Kb\ in \MEG\ (EW = \EWNVIKbMEGWARM\ m\AA), arising from this gas, is detected.

%
%

    \section{Conclusion} \label{sec:conclusions} 

        Our detections confirm the presence of the \hot\ component of the Milky Way's CGM coexisting with a \warm\ and \warmhot\ phases. The characteristic high temperature at \logT\ $\sim$ 7.5 of this component where \SiXIV\ and \SXVI\ are detected is further supported by the non-detection of \NeX, since \Ne\ is expected to be almost completley ionized into \NeXI\ at those temperatures, and its abundance is not that high as that for Oxygen. This result also excludes the presence of high quantities of gas in the CGM with temperatures \logT\ over $\sim$ 6.5 and below $\sim$ 7.5. This \hot\ component is permeating the Milky Way's CGM, and points to be homogeneous in temperature and inhomogeneous in column density.

        In our analysis of \citealp{LaraDI2023}, we inspected for each sightline the presence of prominent absorption lines at $z=0$. None of these individual sightlines present these prominent features. {Moreover, we conducted simulations to evaluate whether the observed detections could arise from only a few sightlines. We determined the likelihood of encountering, in an N-sightlines stacked spectrum, individual spectra presenting an absorption line at least N times stronger than that detected in the stacked spectrum. Our analysis indicates that after stacking 40 spectra, this probability diminishes to a negligible value.} Therefore, our results point to a hot phase being widespread in the halo - since the final stacked spectra is the result of the contribution from many sightlines.

        {In order to be consistent with the DM obtained from FRBs in the literature, the hot CGM gas observed in the halo is either enriched in Oxygen, Silicon, and Sulfur, or has metallicity of at least 4.4 times solar value, or a combination of both. This, along with the sub-solar \Fe/\Oxygen\ found for this component, and that found in the ISM in other works, are key for understanding the heating, cooling and mixing mechanisms within the CGM. In particular, our results provide tantalizing suggestions that the super-virial component of the CGM is the result of feedback from winds arising in the Galaxy.}
        
%
%

   \section*{Acknowledgments} 

        The authors of this work thank the anonymous referee for his/her valuable comments. L.D.I. acknowledges support from CONACYT through the PhD scholarship grant 760672. Y.K. acknowledges support from grant DGAPA PAPIIT 102023. S.M. is grateful for the grant provided by the National Aeronautics and Space Administration through Chandra Award Number AR0-21016X issued by the Chandra X-ray Center, which is operated by the Smithsonian Astrophysical Observatory for and on behalf of the National Aeronautics Space Administration under contract NAS8-03060. S.M. is also grateful for the NASA ADAP grant 80NSSC22K1121. S.D. acknowledges support from the KIPAC Fellowship of Kavli Institute for Particle Astrophysics and Cosmology Stanford University. 
        
%
%

    \section*{Data Availability}

        {All data used in this article are available online via the Chandra Data Archive: https://cda.harvard.edu/chaser/}

%
%

    \bibliographystyle{mnras}
    \bibliography{bibliography} 

%
%

    \bsp	
    \label{lastpage}
    \end{document}